\documentclass[12pt,preprint]{aastex}

\shorttitle{Dynamical Masses of Young Stars}
\shortauthors{Schaefer et al.}

\begin{document}

\title{Dynamical Mass Estimates for Incomplete Orbits: \\
	Young Multiple Stars in Taurus and Ophiuchus}

\author{G. H. Schaefer\altaffilmark{1,2}, M. Simon\altaffilmark{2},  T. L. Beck\altaffilmark{3}, E. Nelan\altaffilmark{1}, and L. Prato\altaffilmark{4}}

\altaffiltext{1}{Space Telescope Science Institute, 3700 San Martin Drive, Baltimore, MD 21218; gschaefer@stsci.edu}
\altaffiltext{2}{Department of Physics \& Astronomy, SUNY Stony Brook, Stony Brook, NY 11794-3800}
\altaffiltext{3}{Gemini Observatory, 670 North A'ohoku Place, Hilo, HI 96720}
\altaffiltext{4}{Lowell Observatory, 1400 West Mars Hill Road, Flagstaff, AZ 86001}

\begin{abstract}

We present recent measurements of the orbital motion in the binaries, DF Tau and ZZ Tau, and in the triples, Elias 12, T Tau, and V853 Oph.  We observed these systems with the Fine Guidance Sensors on the {\it Hubble Space Telescope} and with adaptive optics imaging at the W. M. Keck and Gemini North Observatories.  Based on our measurements and those presented in the literature, we perform preliminary orbital analyses for DF Tau, ZZ Tau, Elias 12 Na-Nb, and T Tau Sa-Sb.  Because the orbital coverage in most of these systems does not yet span a sufficient portion of the orbit, we are not able to find definitive orbit solutions.  By using a Monte Carlo search technique, we explored the orbital parameter space allowed by the current set of data available for each binary.  We constructed weighted distributions for the total mass of the binaries derived from a large sample of possible orbits that fit the data.  These mass distributions show that the total mass is already well-defined.  We compute total mass estimates of 0.78$^{+0.25}_{-0.15} M_\odot$, 0.66$^{+0.15}_{-0.11} M_\odot$, 1.13$^{+0.36}_{-0.09} M_\odot$, and 4.13$^{+1.58}_{-0.97} M_\odot$ for DF Tau, ZZ Tau, Elias 12 Na-Nb, and T Tau Sa-Sb respectively, using a distance of 140 pc.  For Elias 12 Na-Nb, where the orbital coverage spans $\sim 164^\circ$, we compute a preliminary orbit solution with a period of $\sim$ 9-12 years.  By including an earlier lunar occultation measurement, we also find a likely orbit solution for ZZ Tau, with a period of $\sim$~32~years.   With additional measurements to continue mapping the orbits, the derived dynamical masses will be useful in constraining the theoretical tracks of pre-main sequence evolution.

\end{abstract}

\keywords{binaries: visual, stars: pre-main sequence, stars: fundamental parameters}

\section{Introduction}

Determining masses and ages of pre-main sequence (PMS) stars is important for understanding the physical processes of star formation.  For clusters of young stars, these parameters are commonly determined from a star's location relative to evolutionary tracks on a color-magnitude diagram.  However, different sets of theoretical tracks yield discrepant results, particularly for low-mass (M $< 1 M_{\odot}$) young stars, where the predicted masses and ages can differ by factors as large as 2-3 \citep{simon01,simon05}.  These discrepancies result mostly from differences in the choice of atmospheric opacities and the treatment of interior convection.

Calibration of theoretical evolutionary tracks requires comparison with high precision dynamical mass measurements.  Methods used to measure dynamical masses of pre-main sequence stars include: (1) observations of eclipsing double-lined spectroscopic binaries, where analysis of the radial velocity variations and the eclipse light curve yield the individual masses and radii of the component stars \citep{popper87,casey98,covino00,alencar03,stassun04}; (2) mapping the Keplerian rotation of circumstellar disks to determine the mass of the central star \citep{simon00,dutrey03}; and (3) combining spectroscopic and spatially-resolved observations of binary star orbits to determine the component stellar masses and distance to the system \citep{steffen01,boden05}.

Currently, there are only a small number of low-mass PMS stars accessible to each of these techniques.  Observing eclipsing binaries, for instance, requires the orbit be inclined by $\sim 90^{\circ}$ along our line of sight, whereas single stars with rotating circumstellar disks must have disks large enough to detect and resolve with millimeter-wave interferometers.  Moreover, visual and spectroscopic detections of young binaries tend to sample systems distributed in different regions of orbital parameter space.  PMS spectroscopic binaries typically have periods measured on the order of a few years or less \citep{mathieu94}, corresponding to radial velocities amplitudes on the order of tens of km~s$^{-1}$.  Meanwhile, instrumental angular resolution limits the detection of visual binaries with similarly short periods.  At the distance of the Taurus and Ophiuchus star forming regions (140 pc), the resolutions achieved through speckle interferometry and adaptive optics imaging at large 8-10 meter class telescopes or by the {\it Hubble Space Telescope} ({\it HST}) are capable of resolving visual binaries with periods on the order of decades.  The current generation of long-baseline optical/IR interferometers brings visual binary measurements \citep{boden05,schaefer05} further into the orbital parameter space of PMS spectroscopic binaries. 

Through a statistical comparison based on a sample of PMS dynamical mass measurements, only 9 of which are below 1~$M_{\odot}$, \citet{hillenbrand04} showed that for masses above $\sim 1.2 M_{\odot}$, the evolutionary models reproduce the measured masses within 1$\sigma$ on average.  However, for masses below $1.2 M_{\odot}$, most of the evolutionary models systematically underpredict stellar masses by 10-30\%, with the models of \citet{baraffe98} providing the most consistent results.  The goal of the work presented in this paper is to increase the number of low-mass PMS stars with dynamical mass measurements, thereby strengthening the tests of the evolutionary tracks.

In this paper, we present results from our ongoing program \citep{simon96,schaefer03} to map visual orbits of low-mass PMS binaries using the Fine Guidance Sensors on {\it HST} and adaptive optics imaging at the W. M. Keck and the Gemini North Observatories.  We present recent measurements of the separation of the young multiple systems DF Tau, ZZ Tau, Elias 12, T Tau, and V853 Oph.  For ZZ Tau and the close binary (Elias 12 Na-Nb) in the Elias 12 triple, we present preliminary orbit solutions.  For the remainder of the systems, it is not possible to determine a reliable set of orbital elements because our measurements do not yet span a significant fraction of the orbital period.  However, even with the limited orbital coverage, useful mass estimates can be determined based on the observed curvature in the measured orbit.  In this paper, we refine the statistical technique developed in \citet{schaefer03} for determining the total mass of a visual binary when less than half of the orbit is mapped.  We have improved the statistical analysis by incorporating a Monte Carlo search to sample the orbital parameter space and by weighting each orbital solution by its $\chi^2$ probability when analyzing the distributions of masses produced by the search.  We present our analysis of the orbital solutions in \S~\ref{sect.orbit} and discuss the dynamical masses computed from these results in \S~\ref{sect.mass}.  We conclude with a comparison of our dynamical mass estimates with predictions from evolutionary tracks and a brief discussion on the variability observed in the multiple systems.

\section{Observations and Data Reduction}

The stellar properties of the young multiples observed are listed in Table~\ref{tab.targlist}.  The table specifies the target's HBC number, right ascension and declination, V-band magnitude, B-V color, K-band magnitude, spectral type, and luminosity.  If a binary does not have a spectral type identification for each component, then the spectral type of the unresolved system and the total luminosity of the binary are given.  The last column provides an ordered list of references for the values in the table.

\subsection{Fine Guidance Sensor Observations}

The Fine Guidance Sensors (FGS; Nelan et al. 2004) are a set of three optical interferometers on board the {\it HST}.  Each FGS consists of a pair of Koesters prisms that detect the tilt of the incoming wave front in two orthogonal (x and y) directions.  For a given FGS, the normalized difference in counts emerging from each side of the prism, as a function of the tilt of the incoming wavefront, produces the interferometric fringe pattern (S-curve) of the target.  For a binary system, the interferometric response is the linear superposition of two single-star S-curves, scaled by the relative brightness of the components and shifted according to their separation on the sky.

We used FGS1r in its high angular resolution transfer mode to observe our targets in the F853W filter.  Table~\ref{tab.fgs-obs} lists the dates of the FGS measurements presented in this paper.  Each observation was obtained during one {\it HST} orbit and is composed of multiple ($>$10) interferometric scans.  Scan lengths of 1.0$''$ to 1.5$''$ were used to measure the FGS baseline on either side of the binary.  The raw x and y FGS scans were cross-correlated, co-added, and smoothed by piecewise polynomial fits to retain the interferometric signal but remove the effects of the statistical noise of counting photons (see Schaefer et al. 2003; Simon, Holfeltz, \& Taff 1996).

FGS observations of a binary can be deconvolved by modeling the system with the observation of an unresolved single star.  We selected the FGS single-star calibrators SAO 185689 and HD 233877 because their colors (B-V = 1.50 and 1.10, respectively) closely approximate those of the targets.  We used software distributed by the Space Telescope Science Institute (STScI) to analyze the FGS observations of binary systems \citep{nelan02}.  The routine follows a least-squares approach of searching through a grid of separations and brightness ratios for the two components \citep{lattanzi92} and finding the minimum $\chi^2$  between the model and the FGS data.

\begin{figure}
	\scalebox{0.46}{\includegraphics{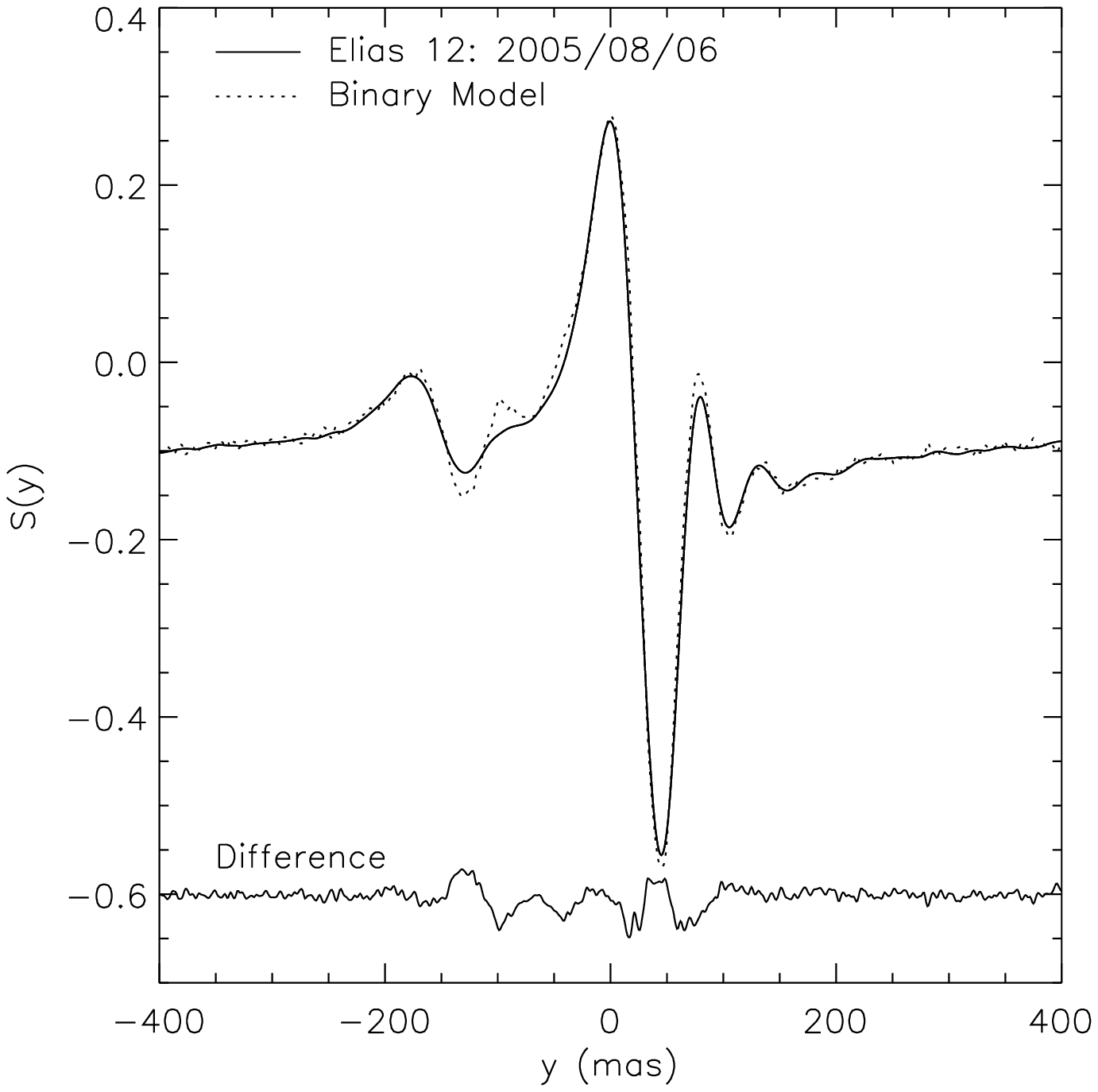}}
	\scalebox{0.46}{\includegraphics{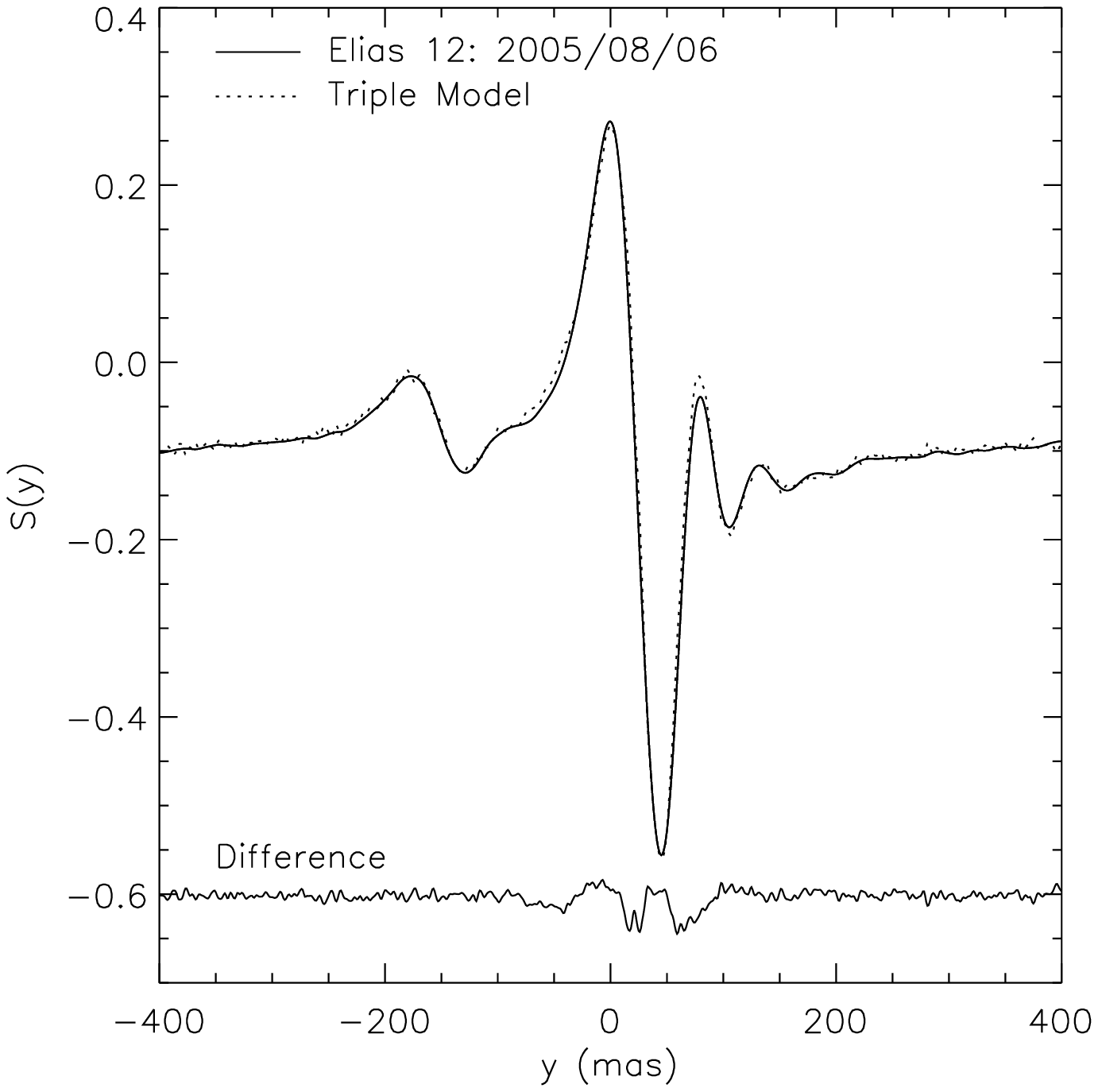}}
	\caption[Modeling of Elias 12 FGS scans]{{\it Left:} Smoothed FGS scan of Elias 12, along the y-axis, on 2005 August 6.  The dotted line shows the best-fit binary model, using SAO 185689 as the PSF.  The residual signature of the close companion can be seen at position of $\sim -100$ mas.   {\it Right:} Elias 12, modeled as a triple system.   The components in the close pair are separated by 30.4 mas along the y-axis of the FGS, with a flux ratio of 0.5.  There is a significant improvement in the residuals of the triple star fit as compared to those of the binary model.}
\label{fig.fgsfit}
\end{figure}

For triple systems (i.e. Elias 12), we developed a program that follows a similar least-squares method.  The routine searches over a grid of positions and brightness ratios for three stars.  As with the STScI software for binaries, the best-fit solution is found from where the $\chi^2$  between the model and the FGS data is at a minimum.  We searched down to step sizes of 0.1 mas in separation and 0.0001 in fractional flux.  When applied to binary systems, the results derived from our program are consistent with those obtained with the software designed at the STScI.  Figure~\ref{fig.fgsfit} shows the model fit for the y-axis of the triple system, Elias 12, on 2005 August 6.  The left panel shows the residuals when Elias 12 is fit as a binary (fitting the close pair as a single source), while the right panel shows the improvement when Elias 12 is fit as a triple.

The uncertainties in the positions and fractional fluxes are determined from the parameter range that causes a variation of 1 from the minimum $\chi^2$.  The largest source of error is dominated by a mismatch between the S-curves of the target and calibrator.  For the purpose of calculating the 1-$\sigma$ uncertainty intervals, we scaled the $\chi^2$ so that the minimum $\chi^2$ is equal to the number of degrees of freedom, $\nu$ (i.e. the reduced $\chi^2_{\nu}$ = $\chi^2/\nu$, is equal to 1).

After the position of each component is measured from the FGS scans, the angular separation, $\rho$, of the stars on the sky is given by
\begin{equation}
\rho = \sqrt{\Delta \theta^2_x + \Delta \theta^2_y}
\label{FGS_sep}
\end{equation}
where $\Delta \theta_x$ and $\Delta \theta_y$ are the projected separations on the x and y axes of the FGS.  The position angle (PA) of the system, measured east of north, is determined from
\begin{equation}
PA = 2 \pi - \phi + \arctan{(\frac{\Delta \theta_y}{\Delta \theta_x})},
\label{FGS_PA}
\end{equation}
where $\phi$ is the spacecraft roll angle \citep{holfeltz95}.

\subsection{Near-IR Adaptive Optics Observations}

We used the adaptive optics (AO) systems at the W. M. Keck and Gemini North Observatories to obtain high resolution images of the multiple systems.  For the hierarchical triples (Elias 12, T Tau, V853 Oph), the wide single component is bright and lies within the isoplanatic patch of the close pair and therefore provides a simultaneous point-spread function (PSF) for measuring the separation and flux ratio of the close pair.  Measuring the separation of the binaries requires using a nearby single star as a PSF reference.  To minimize changes in the AO correction, we selected the single star DN Tau as a PSF because its brightness is similar to that of our targets.  We observed the PSF star immediately before or after the science observations on the targets.

At the W. M. Keck Observatory, we used the near-infrared camera, NIRC2 \citep{wizinowich00}, with the adaptive optics system on the 10-m Keck II Telescope.  The NIRC2 detector is a 1024$\times$1024 Alladin-3 InSb array.  The images were taken with the narrow-field camera which has a plate scale of 0.00994$\pm0.0005''$~pixel$^{-1}$ and an orientation of $0.7^{\circ}\pm0.2^{\circ}$ in position angle (R. Campbell, private comm.).  For each target, we obtained sets of dithered images, using a five-point dither pattern, with a dither offset $\sim 2 ''$.  

We also obtained high resolution images with the near infrared imager, NIRI \citep{hodapp03a,hodapp03b}, using the Altair adaptive optics system \citep{herriot00,veran03} on the 8-meter Gemini North Telescope.  NIRI contains a 1024$\times$1024 Alladin InSb array.  We used it in the f/32 configuration which has a plate scale of 0.0218$''$~pixel$^{-1}$ and an orientation of PA=$0.0^{\circ}\pm0.05^{\circ}$ \citep{beck04}.  For each target, we obtained sets of dithered images, using a five-point dither pattern, with a dither offset $\sim 2 ''$.  

AO images of V853 Oph were also obtained with the slit-viewing camera, SCAM \citep{mclean00}, during an observing run with the NIRSPEC spectrometer on the Keck II telescope.  The SCAM detector is a $256\times256$ HgCdTe array.  When used with the AO system, as we did, it has a plate scale of 0.017$''$ pixel$^{-1}$ \citep{wizinowich03}. 

\begin{figure}
	\scalebox{0.92}{\includegraphics{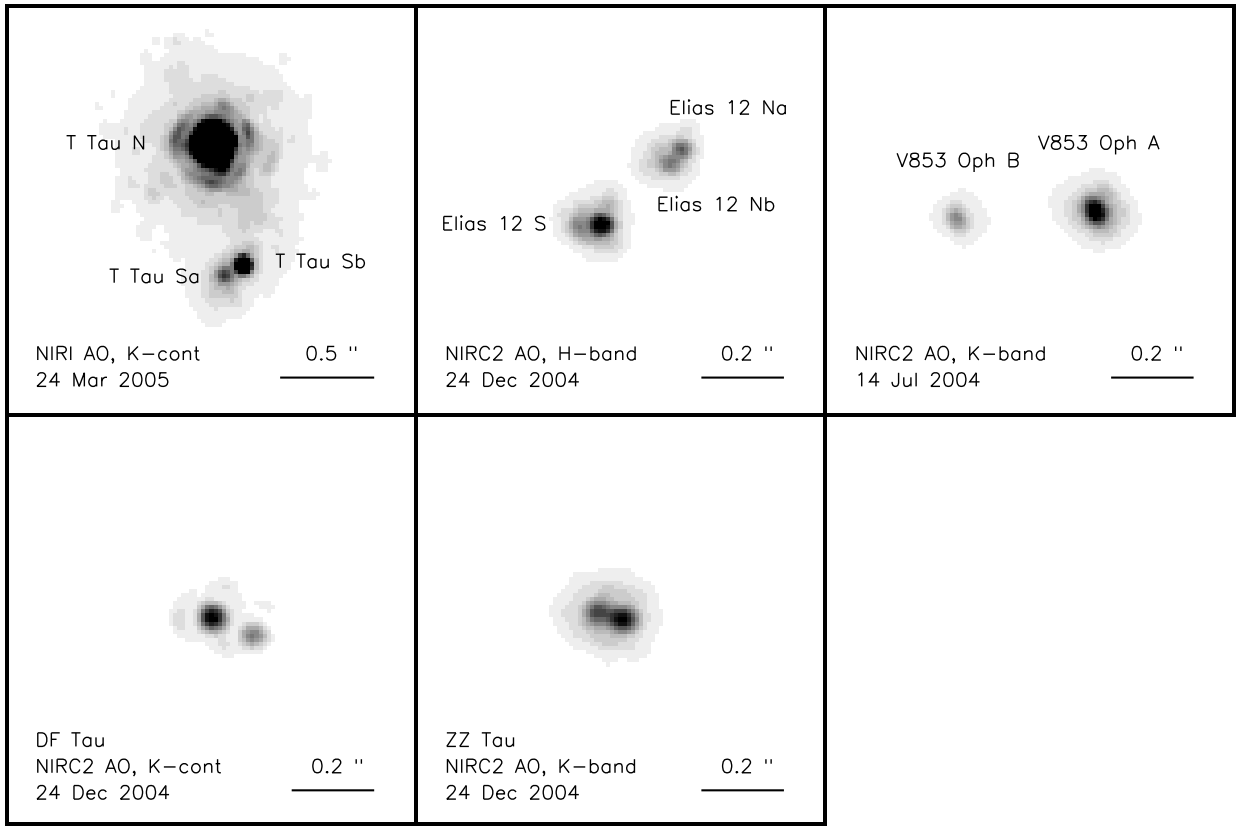}}
	\caption[AO images]{Co-added images of T Tau, Elias 12, V853 Oph, DF Tau, and ZZ Tau taken with adaptive optics imaging using NIRC2 on the Keck II Telescope and NIRI on the Gemini North Telescope.  The camera, filter, and date of observation are listed in the panels for each image.}
\label{fig.aoimages}
\end{figure}

Table~\ref{tab.ao-obs} lists for each multiple system, the date of the AO observations, the instrument and filters used, AO correction rates, integration times, number of co-added exposures per image, and total number of images obtained.  During the observations, the brightest component in each system was used as the AO guide star.  The Strehl ratios, estimated by comparison with an Airy function, ranged from 0.1 to 0.3, depending on the seeing and the brightness of the target.  The images were flatfielded using dark-subtracted, median filtered flat fields obtained on the nights of observation.  Pairs of dithered exposures of the objects were subtracted to remove the background.  Figure~\ref{fig.aoimages} shows co-added AO images for T Tau, Elias 12, V853 Oph, DF Tau, and ZZ Tau.  The close companions in Elias 12 and T Tau are easy to recognize.  In the image of T Tau, T Tau Sa appears fainter than Sb (see Beck et al. 2004 and Duch\'{e}ne et al. 2005 for a detailed discussion of the photometric variability in the T Tau system).  The close pair in V853 Oph A (the brighter component) does not appear to be resolved.

For each of the multiple systems, we extracted subarrays from the image centered on the close binaries, with widths of 0.12$''$-0.30$''$.  For Elias 12, T Tau, and V853 Oph, we used the wide components (Elias 12S, T Tau N, and V853 Oph B, respectively) as the PSF.  For DF Tau, and ZZ Tau we used separate observations of the single star, DN Tau, as a PSF reference.  Since the NIRC2 observations of Elias 12 on 2004 December 24 were saturated on Elias 12N, we used an observation of DN Tau as the PSF for these images also.  For all of the images, we used the PSFs to construct models of the close binaries by searching through a grid of separations and flux ratios. We used the IDL INTERPOLATE procedure to shift the PSF by sub-pixel intervals using cubic convolution interpolation.  To remove any excess light of the primary star from the extracted sub-array of the close pair, we followed a least-squares approach to fit the background as a slanting plane.  We determined the position and flux ratio of the companions in the close binary from where the $\chi^2$ between the model and the observation reached a minimum.  The uncertainties were determined by analyzing multiple images individually and computing the standard deviation.  

\section{Results}

Table~\ref{tab.sepPA} lists the results from the most recent FGS and AO measurements for DF Tau, ZZ Tau, Elias 12, T Tau, and V853 Oph.  The separation and position angle, measured east of north, are given for each epoch of observation.  The last column specifies the instrument used to obtain the measurement.  For the triples, Elias 12 and T Tau, we report the separation measured for each of the wide pairings (Elias 12 S-Na, Elias 12 S-Nb, T Tau N-Sa, and T Tau N-Sb), as well as the relative separation of the close pairs (Elias 12 Na-Nb and T Tau Sa-Sb).  Because of improvements in the method of data analysis and error estimation, we report updated results for the FGS observations of Elias 12 and the AO observations of Elias 12 and T Tau that were presented previously in \citet{schaefer03} and \citet{beck04}.  We attempted to model the observations of V853 Oph as a triple system, however, it was not possible to obtain a reliable set of measurements for the close pair.  As a result, we list in Table~\ref{tab.sepPA} only the relative position of the wide pair, V853 Oph A-B, fitting the close pair, V853 Oph A, as an unresolved source.  

The FGS records the number of counts detected by the photomultipliers (PMT) every 25 msec.  We used the flux ratios measured from the FGS scans to convert the average number of counts per 25 msec received from the target to the PMT counts contributed by each stellar component.  These values are given in Table~\ref{tab.fgscounts} for DF Tau, ZZ Tau, and Elias 12.  Table~\ref{tab.aoflux} lists the flux ratios measured from the adaptive optics observations.    For the triples, we list the flux ratio of each component relative to the widely separated member, as well as the flux ratio between the two components in the close pair.  For the binaries, we list only the flux ratio between the two components.  For each date of observation, we indicate the filter used for the listed flux ratios.  Technical specifications of the filters are given in the instrument manuals at the observatory web sites.  Since the flux of T Tau N has remained constant in recent years, with an average magnitude of $5.53\pm 0.03$ at K \citep{beck04}, we convert the relative fluxes of T Tau Sa and Sb into K-band magnitudes and present these results in Table~\ref{tab.ttaumag}.

\section{Computation of Orbital Parameters}
\label{sect.orbit}

Based on the cumulative orbital motion measured for each of the binaries, we calculated orbital fits by following the three-dimensional grid search procedure of \citet{hartkopf89} and \citet{mason99}.  We explored the parameter space by stepping through ranges of the period, time of periastron passage, and eccentricity $(P,T,e)$.  The program locates the orbital solution for which the $\chi^2$ between the measured and calculated positions at the times of observation is at a minimum.

Input to the orbit fitting program includes a set of starting values $(P,T,e)$, the corresponding step sizes and search ranges for each of the three parameters, and the measured separation and position angle at the times of observation.  At each step through the three-dimensional grid, a least-squares fit is perfomed \citep{hartkopf89} to determine the Thiele-Innes elements ($A,B,F,G$). The angular semi-major axis ($a$), inclination ($i$), position angle of the line of nodes ($\Omega$), and the angle between the node and periastron ($\omega$) are computed from the Thiele-Innes elements (i.e. Couteau 1981).  The data are weighted by their respective measurement errors and the $\chi^2$ between the calculated $(x_c,y_c)$ and observed positions $(x_o,y_o)$ are computed from
\begin{equation}
\chi^2 = \sum \left(\frac{(x_o - x_c)^2}{\sigma_x^2} + \frac{(y_o - y_c)^2}{\sigma_y^2}\right) 
\end{equation}
where the errors, $\sigma_x$ and $\sigma_y$, are propagated from the measurement uncertainties listed in Table~\ref{tab.sepPA}.  The set of  parameters that produces a minimum in the $\chi^2$ surface is retained as the best-fit orbital solution.

Formal errors in the orbital elements can be derived using the general matrix method for multiple regression \citep{bevington92}.  These error estimates involve linearizing the equations of orbital motion through a Taylor expansion terminated after the first order. 
For orbits that are poorly determined, higher order terms may become important.  For such cases, it is necessary to examine the shape of the $\chi^2$ surface.  In such cases, the precision of the computed orbital elements can be evaluated from the distribution of orbital elements that produces a variation of one from the minimum $\chi^2$.

For binaries where the orbital coverage spans less than 180$^{\circ}$, it is not possible to determine a reliable set of orbital parameters \citep{hartkopf01}.  This is the case for most of the systems presented in this paper.  In order to explore the $\chi^2$ surface for each of the binary orbits, we performed a Monte Carlo search by modifying the three-dimensional grid search procedure so that the values of $P$, $T$, and $e$ on each pass through the program are chosen at random.  If a set of orbital parameters yields a $\chi^2$ value that is within $\Delta\chi^2 = 1$ of the minimum, the solution is stored.  After a specified number of possible solutions are found (N=10,000), the results are plotted and examined.  The projection of the $\Delta\chi^2 = 1$ surface onto each of the orbital parameters corresponds to the 1-$\sigma$ uncertainty interval, assuming that the parameters are independent \citep[Ch.15]{press92}.  During this procedure, we scale the $\chi^2$ values so that $\chi^2_\nu = 1$ at the minimum.  

Because of the limited orbital coverage for the binaries, the $\chi^2$ surfaces turn out to be broad and flat in the region of the minimum.
Table~\ref{tab.orbpar} lists the 1-$\sigma$ uncertainty intervals for the orbital parameters of DF Tau, ZZ Tau, Elias 12 Na-Nb, and T Tau Sa-Sb.  Upper search ranges of P=500 years and e=0.99 were used when exploring the parameter space.    If the $\Delta\chi^2=1$ interval extends beyond these search ranges, only the 1-$\sigma$ lower limit is quoted.  The results for the individual systems are described in more detail below.  In Section~\ref{sect.mass}, we discuss how the broad range of allowable parameters influences estimates of the total system mass.

\subsection{DF Tau}

DF Tau is a classical T Tauri star \citep{HBC} that was first resolved as a binary during a lunar occultation survey \citep{chen90}.  From angularly resolved {\it HST} STIS spectra, \citet{hartigan03} identified the spectral types of the primary and secondary as M2.0 and M2.5, respectively.  Since its discovery, the orbital motion of the DF Tau binary has been mapped through speckle interferometry \citep{ghez93,ghez95,thiebaut95,balega02,balega04,shakhovskoj06}, {\it HST} imaging \citep{white01}, and FGS interferometery \citep{simon96,schaefer03}.  Preliminary orbit solutions were computed by \citet{thiebaut95}, \citet{tamazian02}, and \citet{shakhovskoj06}, with periods ranging from 74 to 92 years.  However, the reliability of these results is limited by the sparse orbital coverage.  As illustrated by \citet{schaefer03}, a large range of orbital solutions can fit the available data.

\begin{figure}
   \scalebox{0.9}{\includegraphics{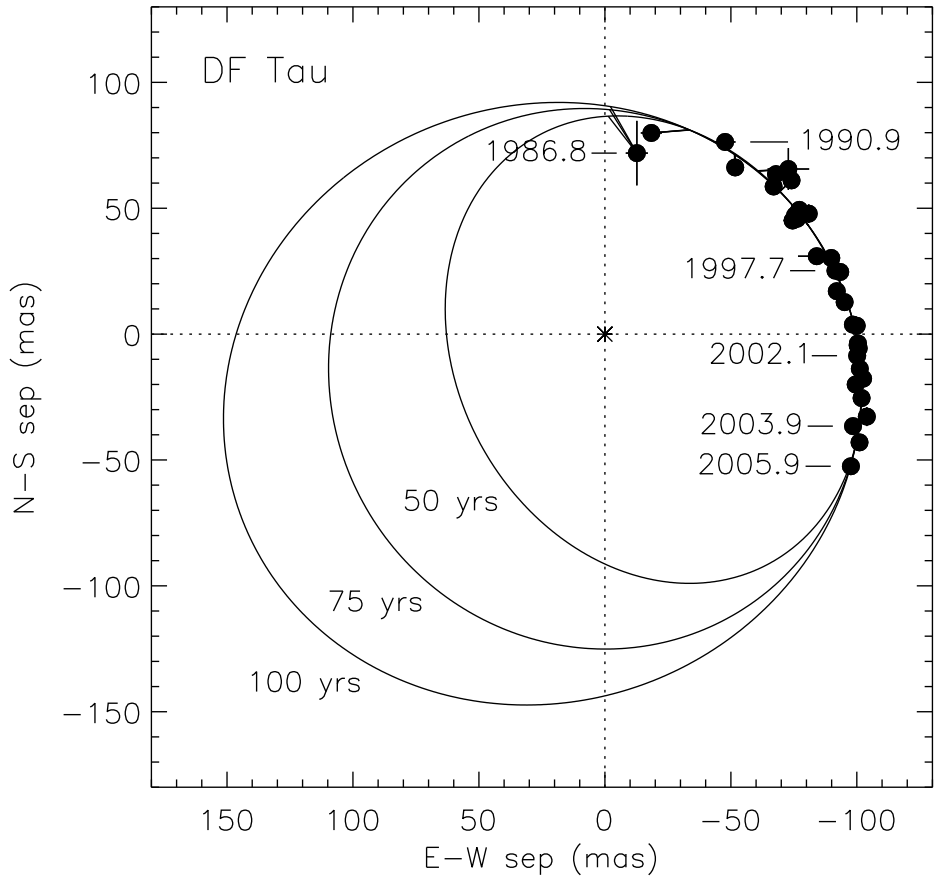}}
   \caption[DF Tau orbital motion]{Orbital motion of DF Tau B (secondary) relative to DF Tau A (primary).  The location of the primary star is marked by an asterisk.  Examples of three possible orbital solutions for DF Tau at periods of 50, 75, and 100 years are overplotted.}
\label{fig.dforb}
\end{figure}

Figure~\ref{fig.dforb} shows the orbital motion of DF Tau, using the cumulative astrometric data in the literature (referenced above) and our recent measurements presented in Table~\ref{tab.sepPA}.  The orbital coverage spans $\sim 110^{\circ}$.  By exploring the $\chi^2$ surface using the Monte Carlo grid search technique, we found a minimum located at $P \sim 95.8$ years.  However, since the 1-$\sigma$ interval extends from a period of 37 years to periods greater than 500 years, the precise orbital period is not definitive.  Figure~\ref{fig.dforb} shows three examples of possible orbital solutions for DF Tau at periods of 50, 75, and 100 years, which agree within $\Delta \chi^2 < 0.4$.  

\begin{figure}
	\scalebox{0.9}{\includegraphics{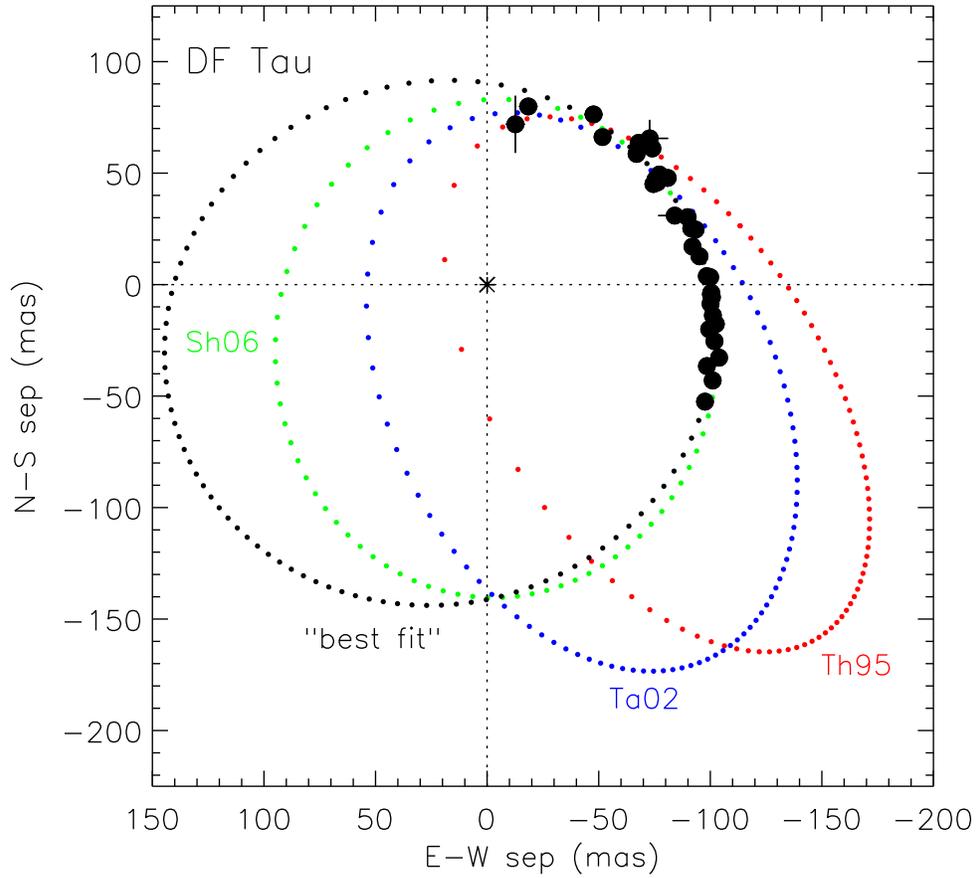}}
	\caption[Previously published orbits of DF Tau]{Previously published orbits of DF Tau computed by Thi\'{e}baut et al. (1995, Th95 - red), Tamazian et al. (2002, Ta02 - blue), and Shakhovskoj et al. (2006, SH06 - green). We also include the orbital solution that we find at the minimum of the $\chi^2$ surface through our Monte Carlo search (``best fit'' - black).  The orbits are plotted at yearly intervals.}
\label{fig.dfpub}
\end{figure}

Figure~\ref{fig.dfpub} compares the current set of measurements for DF Tau to the orbital solutions computed by \citet{thiebaut95}, \citet{tamazian02}, and \citet{shakhovskoj06}.  We also plot the 95.8 year orbit that we found at the minimum of the $\chi^2$ surface through our Monte Carlo search.  It is clear that the solutions of \citet{thiebaut95} and \citet{tamazian02} are no longer consistent with the current measurements.  It is likely that the orbit of \citet{shakhovskoj06} and our best-fitting solution will require substantial revision in the years to come.   

The orbital calculations for DF Tau indicate that the system has not been observed with sufficient coverage to determine a definitive set of orbital parameters.  This is consistent with its designation as a grade 4 orbit in The Sixth Catalog of Orbits of Visual Binary Stars \citep{hartkopf06}.  This grade assignment indicates that any orbital determination should be considered preliminary, where substantial revision may be required for the individual elements.  This agrees with the results presented here.

\subsection{ZZ Tau}

Observed as an unresolved system, ZZ Tau  is classified as a classical T Tauri star with a spectral type of M3 \citep{HBC}.  A lunar occultation observation \citep{simon95} revealed the binary nature of the system.  The orbital motion of ZZ Tau has been monitored with the FGS since 1994 \citep{simon96, schaefer03}.  A preliminary orbit analysis by \citet{schaefer03} indicated that a large range of orbits fit the available data.

\begin{figure}
   \scalebox{0.9}{\includegraphics{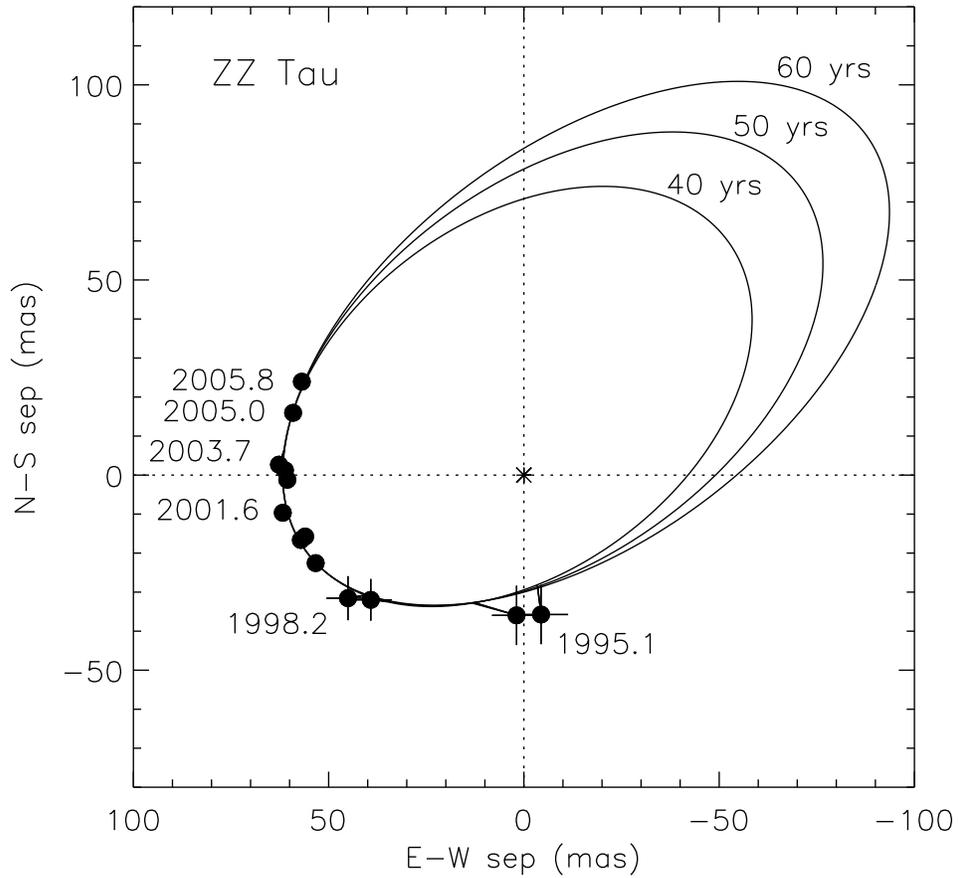}}
   \caption[ZZ Tau orbital motion]{Orbital motion of ZZ Tau B (secondary) relative to ZZ Tau A (primary) as measured by our FGS and AO observations.  The location of the primary star is marked by an asterisk.  Examples of three possible orbital solutions for ZZ Tau, at periods of 40, 50, and 60 years, are overplotted.}
\label{fig.zzorb}
\end{figure}

Figure~\ref{fig.zzorb} shows the orbital motion measured for ZZ Tau, based on the cumulative set of FGS and AO measurements (see references above and Table~\ref{tab.sepPA}).  The orbital coverage spans $\sim$ 120$^{\circ}$.  Three examples of possible orbital solutions, at periods of 40, 50, and 60 years which agree within $\Delta \chi^2 < 0.3$, are overplotted in the figure.  The earliest two FGS separation measurements of ZZ Tau are the most discrepant from the orbital fits.  In these observations, the FGS was oriented so that the binary remained unresolved along the x-axis and therefore the positions are derived from the combination of consecutive observations at different {\it HST} roll angles (see Schaefer et al. 2003).

The Monte Carlo grid search technique reveals a nearly flat $\chi^2$ surface out to the search limit of P=500 yrs.  The broad 1-$\sigma$ confidence interval reaches a shallow minimum at a period of $\sim$ 180 years, however, a large concentration of solutions (30\%) exist with periods between 30 to 100 years.  The shallowness of the minimum is caused by limited coverage and sampling of the orbital measurements.  It is therefore not possible to determine a definitive set of orbital elements for ZZ Tau based on the FGS and AO data alone.  

\begin{figure}
   \scalebox{0.9}{\includegraphics{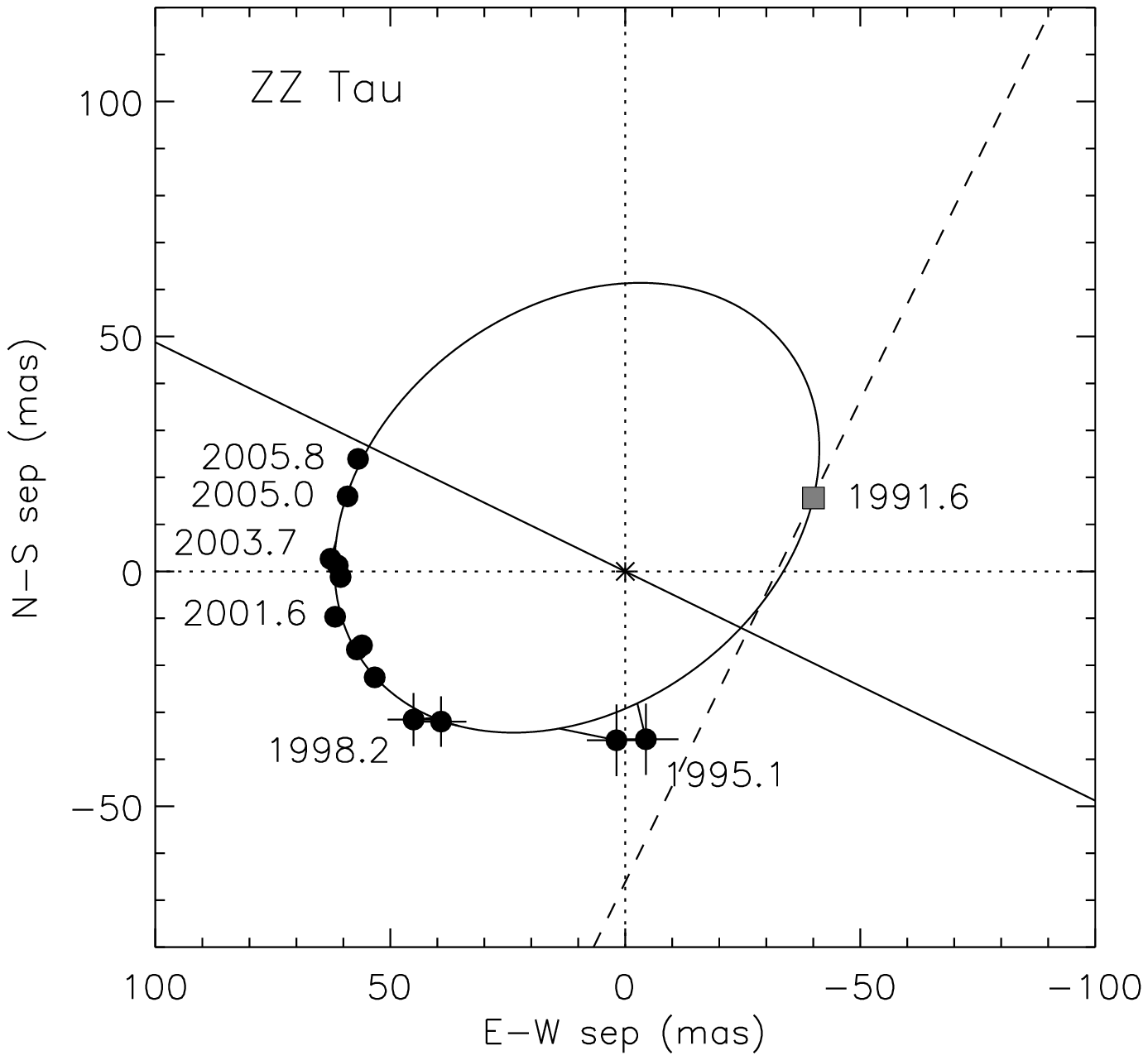}}
   \caption[ZZ Tau lunar occultation]{Same as Figure~\ref{fig.zzorb}, but includes the projected lunar occultation measurement of ZZ Tau from 1991.6 \citep{simon95}.  The solid line shows the direction of the lunar occultation along a position angle of 244$^{\circ}$.  The dashed line shows the projected separation of 29 mas measured along this direction. Overplotted is the best fit orbit (P=31.5 yrs) based on the FGS, AO, and LO measurements. The position in the orbit at the time of the LO is marked by the shaded square.}
\label{fig.zzlunar}
\end{figure}

To improve the orbital coverage of ZZ Tau, we added the lunar occultation measurement observed by \citet{simon95} to our dataset.  During the occultation of ZZ Tau in 1991.6, a projected separation of 29 mas was measured along a position angle of 244$^{\circ}$.  The orientation of the occultation is indicated in Figure~\ref{fig.zzlunar}.  We used a standard Newton Raphson approach to compute a joint fit to the FGS, AO, and lunar occultation (LO) data.  Table~\ref{tab.zztauorb} lists the best fit orbital parameters, the $\chi^2$ of the fit, and the number of degrees of freedom, $\nu$.  Figure~\ref{fig.zzlunar} shows a plot of the best fit orbit.  Based on the 31.5~year period, the addition of the LO measurement increases the orbital coverage by $\sim 100^\circ$.  Continued FGS and AO observations over the next two decades should provide complete coverage across the ZZ Tau orbit.

\subsection{Elias 12}

Elias 12 (HBC 404, V807 Tau) is a hierarchical triple.  The primary, Elias 12S, is a classical T Tauri star \citep{HBC} with a spectral type of K5 \citep{hartigan03}.  The close pair, Elias 12N, is located $\sim$ 300 mas to the north of Elias 12S and is currently separated by $\sim$ 45 mas.  Observed as an unresolved system, Elias 12N, has a spectral type of M2.0 \citep{hartigan03}.  The orbital motion of the wide components, Elias 12 S-N, with the close pair unresolved, was  measured by \citet{ghez93}, \citet{leinert93}, and \citet{simon96}.  \citet{simon95} resolved the projected separation of the close pair along the direction of a lunar occultation.  Our FGS and adaptive optics observations to map the motion as a triple system began in 1999 \citep{schaefer03}.

\begin{figure}
   \scalebox{0.9}{\includegraphics{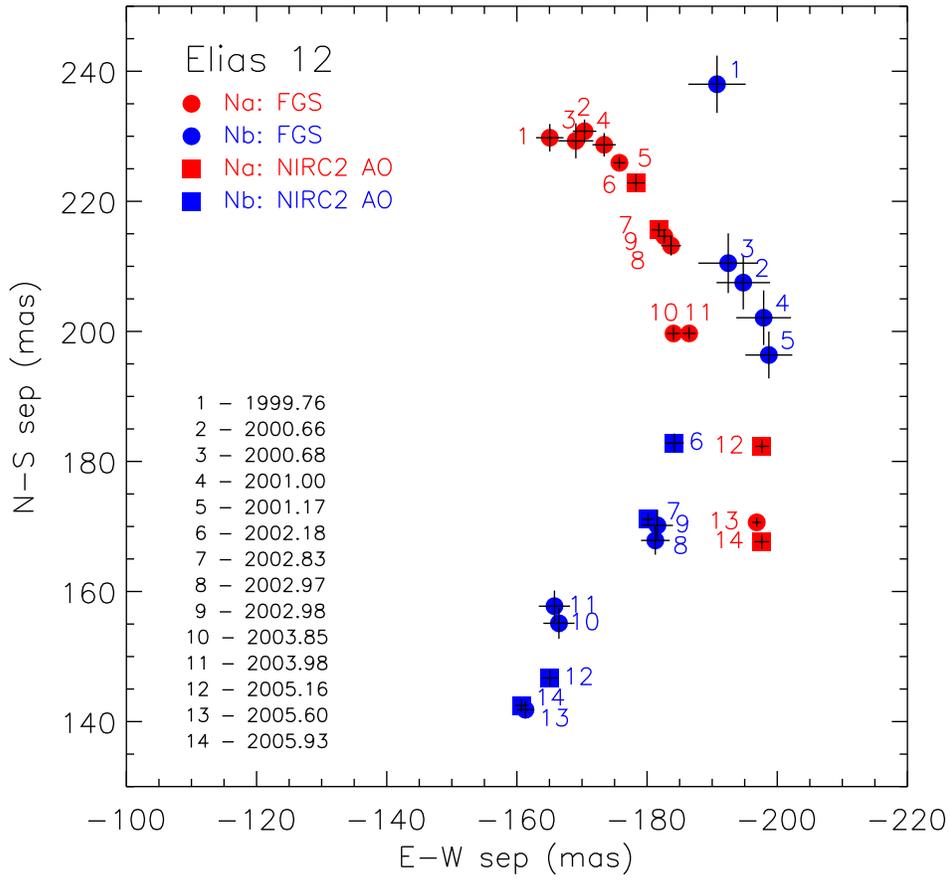}}
   \caption[Elias 12 N-S position measurements]{Orbital motion in the Elias 12 hierarchical triple.  The positions of Elias 12 Na (red symbols) and Nb (blue symbols) are plotted relative to Elias 12S (located at 0,0).}
\label{fig.elwide}
\end{figure}

\begin{figure}
   \scalebox{0.9}{\includegraphics{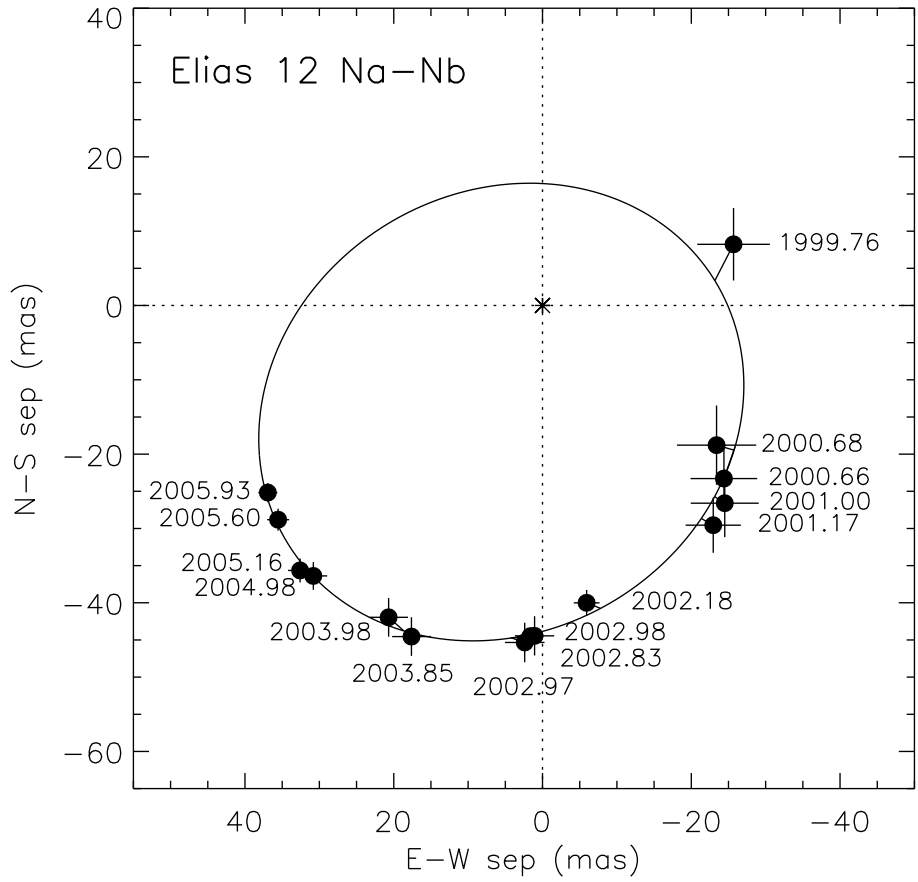}}
   \caption[Elias 12 Na-Nb orbital motion]{Orbital motion of Elias 12 Nb relative to Na (marked by an asterisk). Overplotted is the best fit orbit (P=9.4 yrs) based on the FGS and AO measurements alone.}
\label{fig.elorb}
\end{figure}

Figure~\ref{fig.elwide} shows the motion of Elias 12 Na and Nb as measured relative to the wide component, Elias 12 S.  Figure~\ref{fig.elorb} shows the motion observed for the close binary, where the position of Elias 12 Nb is plotted relative to Elias 12 Na.  Over the 6 years of resolved FGS and AO observations, the close pair has moved $\sim 164^{\circ}$ relative to one another.  Since these measurements span nearly half of the orbit, a preliminary orbit solution can be computed.  Table~\ref{tab.el12orb} lists the best fit orbital parameters computed through a standard Newton Raphson method based on the FGS and AO data.  The $\chi^2$ of the fit and number of degrees of freedom, $\nu$, are listed in the table.  The orbit, with a period of 9.4 years, is overplotted in Figure~\ref{fig.elorb}.  The orbital parameters and uncertainties from the formal fit are similar to those determined from the Monte Carlo search technique, which are presented in Table~\ref{tab.orbpar}.  

\begin{figure}
   \scalebox{0.54}{\includegraphics{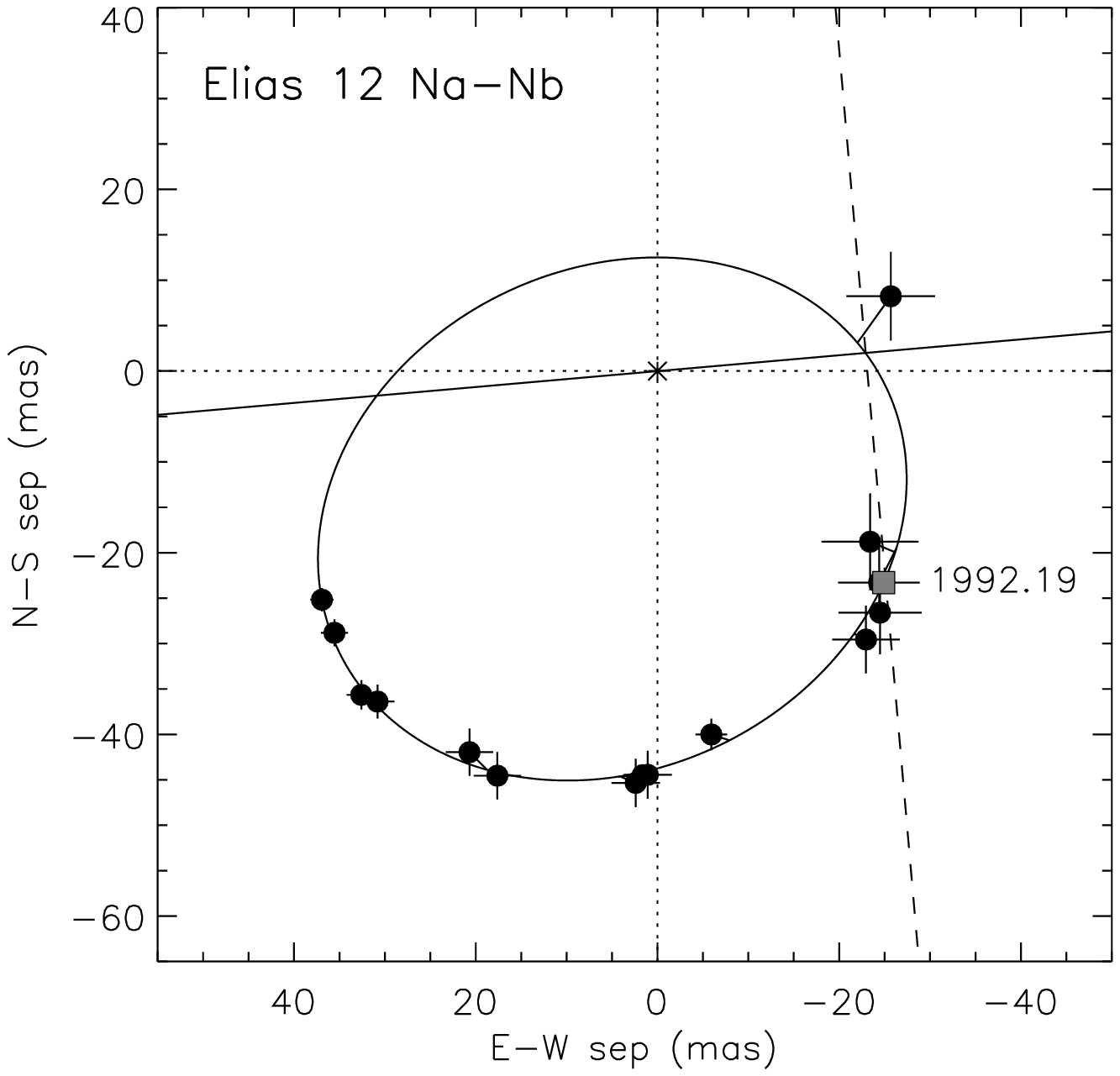}}
   \scalebox{0.54}{\includegraphics{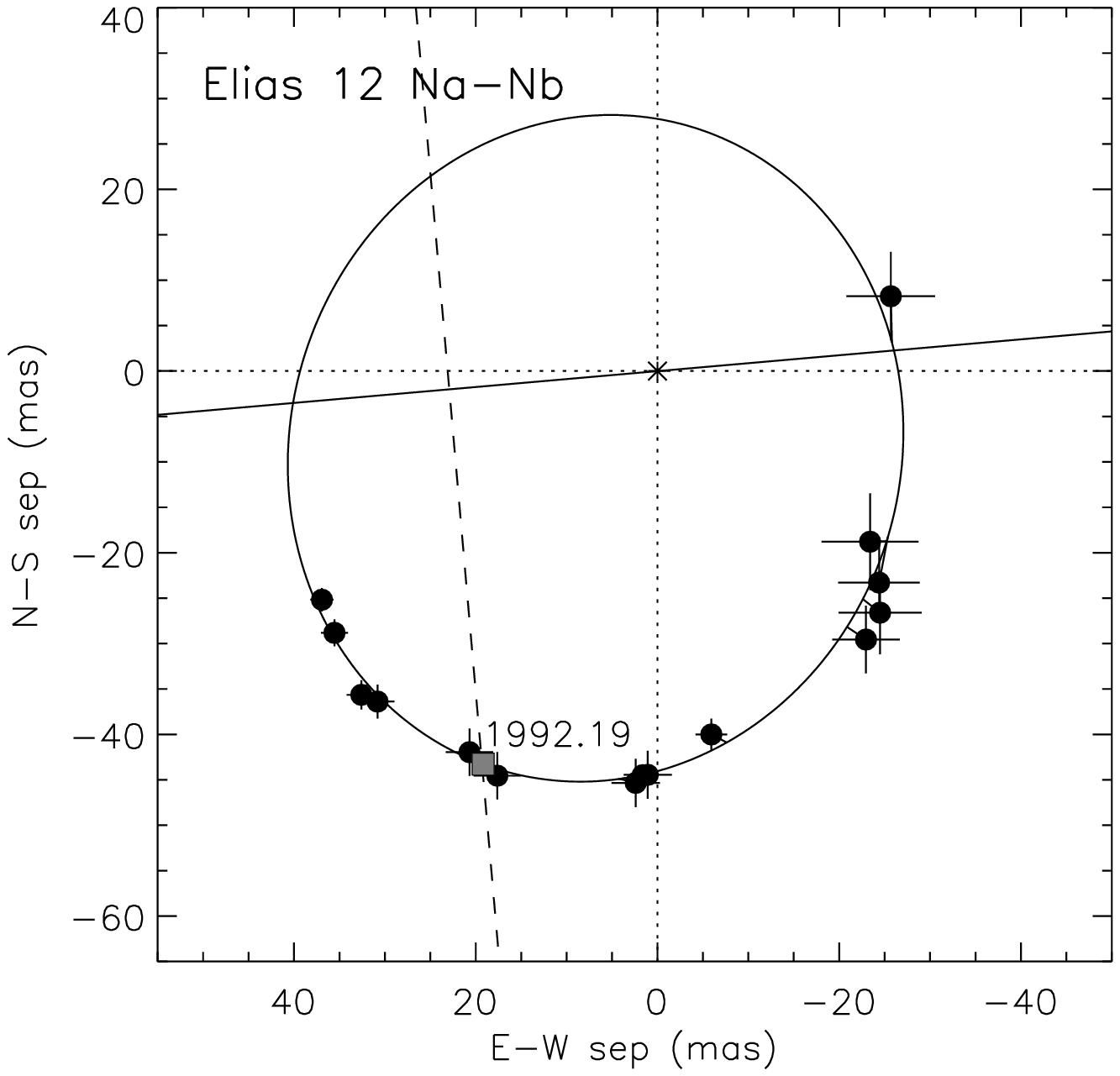}}
   \caption[Elias 12 Lunar Occultation]{{\it Left:} Same as Figure~\ref{fig.elorb}, but includes the projected lunar occultation measurement of Elias 12 from 1992.2 \citep{simon95}.  The solid line shows the direction of the lunar occultation along a position angle of 95$^{\circ}$.  The dashed line shows the projected separation of 23 mas measured along this direction. Overplotted is the best fit orbit (P=8.65 yrs) based on the FGS, AO, and LO measurements. The position in the orbit at the time of the LO is marked by the shaded square.  Adding the LO observation extends the orbital coverage to span more than one cycle of the period.  {\it Right:} Same as the left panel but showing the alternative possibility for the projected separation (rotated by 180$^\circ$).  Overplotted is the best fit orbit (P=11.85 yrs) based on the FGS, AO, and LO measurements.}
\label{fig.ellunar}
\end{figure}

To investigate the reliability of the orbit fit based on the FGS and AO data, we compared our results to the projected separation of $23$~mas measured during the lunar occultation observation along a position angle of 95$^{\circ}$ in 1992.2 \citep{simon95}.  Adding this observation extends the orbital coverage to span more than one full cycle of the computed period.  Because the lunar occultation observation measured equal fluxes for Elias 12 Na and Nb, there is an ambiguity of 180$^{\circ}$ in the projected separation.  The two possible orientations are shown as dashed lines in the panels of Figure~\ref{fig.ellunar}.  For each of these two possiblilities, we computed a joint fit to the FGS, AO, and lunar occultation (LO) data.  The orbital parameters derived for each fit are listed in Table~\ref{tab.el12orb} and overplotted in Figure~\ref{fig.ellunar}.  While the unscaled $\chi^2$ for the P=8.65 year orbit is lower than that for the P=11.85 year by $\Delta\chi^2 = 0.65$, the derived total mass of  $\sim 1.0~M_\odot$ for the 11.85 year orbit is more consistent with the M2.0 spectral type of Elias 12N than the $\sim 2.2~M_\odot$ value for the shorter period.  These values assume a distance of 140 pc, the average distance to the Taurus star forming region \citep{kenyon94}.

The three orbit solutions listed in Table~\ref{tab.el12orb} show that there remains some uncertainty in the orbital parameters.  However, with an orbital period of 9-12 years, a reliable orbit solution will be forthcoming over the next few years.  We will continue to measure the orbital motion of Elias 12 in our ongoing observing programs.  Additionally, since Elias 12 S provides a reference to measure the motion of the close pair about their center of mass, this could allow us to calculate the mass ratio of Elias 12 Na and Nb.  Combined with the total mass from the visual orbit and an estimate of the distance, this would provide the individual masses of each component.

\subsection{T Tau}

T Tau, the prototypical young star \citep{joy45,ambartsumian47,ambartsumian49}, is a hierarchical triple located in the Taurus star forming region.  The visible light primary, T Tau N, has a spectral type of K0 \citep{basri90}.  The infrared companion, T Tau S, located $\sim 0.7''$ to the south of T Tau N, was discovered by \citet{dyck82}.  Modeling of its spectral energy distribution suggests that T Tau S is the most luminous component, having a mass of $\sim 2.5 M_\odot$ and a large visual extinction from circumstellar material \citep{koresko97}.  T Tau S was found to have a close companion at a projected separation of $\sim 0.05''$ through speckle interferometry \citep{koresko00}.  Spatially resolved spectra indicate an intermediate-to-early spectral type for the infrared companion, T Tau Sa, and a spectral type of M1 for the close companion, T Tau Sb \citep{duchene02,duchene05}. T Tau Sa and T Tau Sb remain optically invisible, down to a limiting magnitude of $\sim$ 19.6 \citep{stapelfeldt98}.  Silicate and water-ice features indicate high-levels of foreground material toward T Tau Sa and Sb \citep{ghez91,beck04}.  Additionally, CO absorption in the spatially resolved spectrum of T Tau Sa suggests the presence of a circumstellar disk viewed nearly edge-on and located within $\sim$ 3 AU of the star \citep{duchene05}.  \citet{beck04} discuss the cumulative measurements of the apparent motion of T Tau S relative to T Tau N.  Since its discovery, the orbital motion of T Tau Sb relative to Sa has been monitored by \citet{koresko00}, \citet{kohler00}, \citet{duchene02}, \citet{furlan03}, \citet{beck04}, \citet{duchene05}, \citet{mayama06}, and \citet{duchene06}\footnote{We do not include the new astrometric measurements reported by \citet{duchene06} in our orbital analysis, but have added the reference for completeness.  The new positions are consistent with the set of measurements used in our analysis and because they span the same interval of time, would not significantly alter the results from our Monte Carlo search technique.}. 

\begin{figure}
   \scalebox{0.9}{\includegraphics{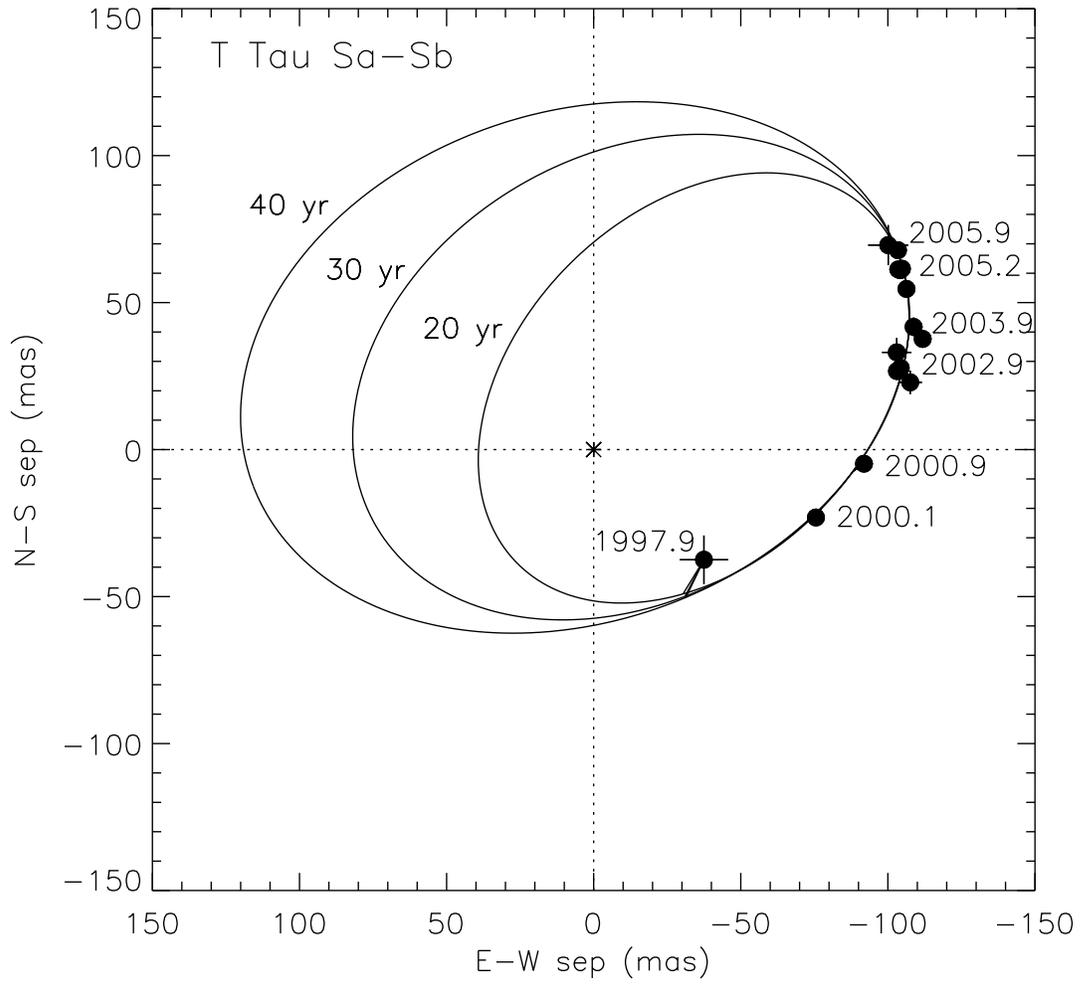}}
   \caption[T Tau Sa-Sb orbital motion]{Orbital motion of T Tau Sb relative to Sa (marked by an asterisk). Examples of three possible orbital solutions, at periods of 20, 30, and 40 years are overplotted.}
\label{fig.ttorb}
\end{figure}

Figure~\ref{fig.ttorb} shows the orbital motion of T Tau Sb relative to T Tau Sa, using the cumulative astrometric data in the literature (see references above) and our recent measurements listed in Table~\ref{tab.sepPA}.  We do not include the measurements reported by \citet{mayama06} because there appears to be a systematic offset of $\sim$ 20 mas in the measured positions they present as compared to the rest of the astrometric data.  Based on the set of infrared measurements, the orbital coverage spans $\sim 80^\circ$ in position angle.  Overplotted in the figure are three examples of possible orbits, at periods of 20, 30, and 40 years, which agree within $\Delta\chi^2 < 0.1$.  The Monte Carlo grid search reveals that the 1-$\sigma$ $\chi^2$ surface is broad and flat and extends out to the search limit of P=500 years.  The minimum in the $\chi^2$ surface occurs at a period of $\sim 20.8$ years.  

The northern and southern components of T Tau are both detected in radio emission.  The southern radio source appears to be associated with the infrared position of T Tau Sb \citep{johnston03, johnston04, loinard03, smith03}.  By combining nearly 20 years of radio observations with the positions measured in the infrared, \citet{johnston04} fit an orbit, with a period of $25.5\pm8.8$ years, to the T Tau Sa-Sb system.  The $\chi^2$ minimum that we find through the Monte Carlo search using only the infrared measurements is consistent with this value.  Our results are also in agreement with the recent astrometric orbital analysis reported by \citet{duchene06}.  The high spatial resolution maps of \citet{johnston04} show that the southern radio source can actually be resolved into two components.  However, the separation between the two radio sources is smaller than that measured for T Tau Sa and Sb in the infrared.  The authors suggest that the radio emission could be due to magnetic reconnection processes in the environment of T Tau S, that is not coincident with either star.  The position of the radio emission is useful for placing constraints on the orbital motion of the T Tau S binary, however, additional infrared and radio observations are necessary to further understand the connection between these measurements.

\subsection{V853 Oph}

As an unresolved system, V853 Oph is classified as a classical T Tauri star \citep{HBC} with a spectral type of M2.5 \citep{bouvier92}.  \citet{ghez93} resolved the wide component of the triple system at a separation of $\sim 400$ mas using speckle interferometry.  In 1992, the primary, V853 Oph A, was discovered to have a close companion at a separation of 13 mas along the projection of a lunar occultation \citep{simon95}, with a measured brightness ratio of 0.83 at K.  Beginning in June 1999, we have monitored the system with the FGS and through AO imaging. 

The diffraction limit of the Keck II Telescope at H is $\sim 40$ mas and that of {\it HST} at V is $\sim 60$ mas.  This makes it difficult to measure reliably the small separation of the close binary in V853 Oph A.  For the AO observations, we used the fainter tertiary, V853 Oph B, as a simultaneous PSF reference. Since V853 Oph B is separated by only $\sim 340$ mas from V853 Oph A and the system lies low in the southern sky at Mauna Kea, the observations require good natural seeing to obtain a well-resolved PSF, even with the use of AO.  For example, during the NIRC2 observations on 2002 March 8 and 2004 June 15, it was not possible to extract the tertiary without including a significant amount of light from the primary.  Resolving the close pair in V853 Oph with the FGS is also complicated by the small separation.  For some FGS scans, there appears to be residual signal remaining after modeling the close pair as an unresolved source.  However, measuring separations smaller than $\sim$ 14 mas on an individual axis of the FGS is not reliable, unless the magnitude difference can be constrained by a well-separated measurement on the other axis \citep{nelan04}.  This was not the case for V853 Oph A.  As a result of these complications, we were not able to derive a reliable set of measurements for the close binary in V853 Oph.

\begin{figure}
   \scalebox{0.9}{\includegraphics{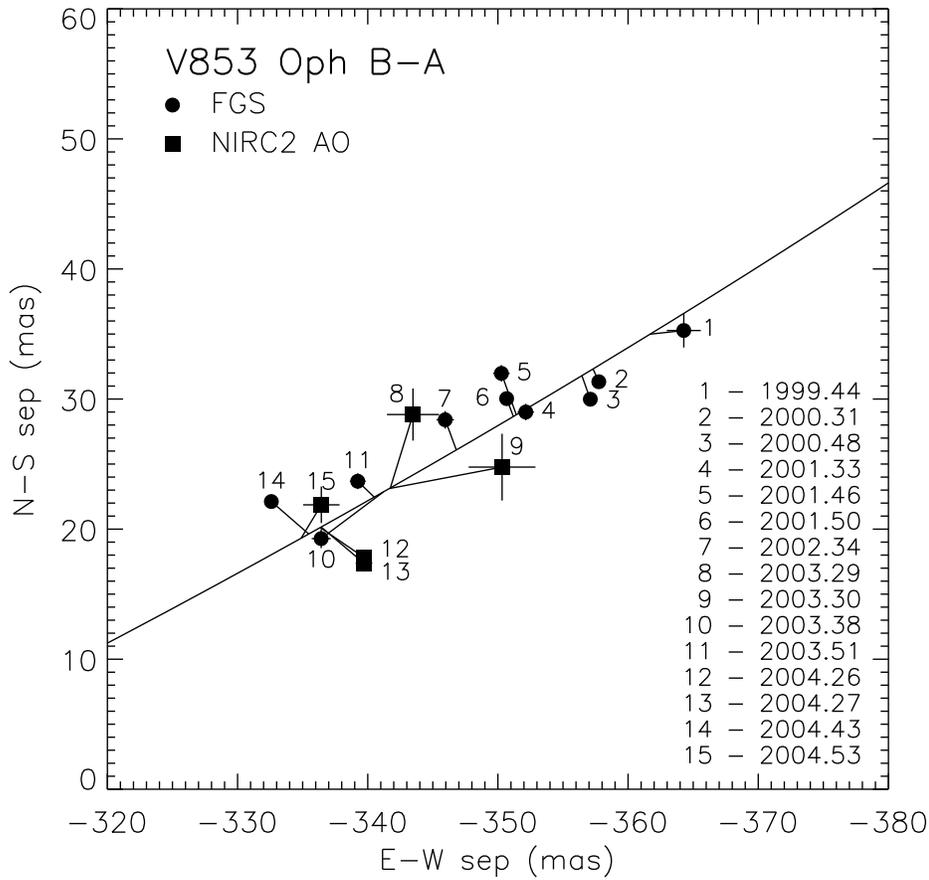}}
   \caption[Orbital motion of V853 Oph B-A]{Motion of V853 Oph A as measured relative to the tertiary, V853 Oph B (located at 0,0).  The positions were determined by modeling the close binary, V853 Oph A, as an unresolved source.  Overplotted is a model orbit with a period of 600 yrs and a total mass of 1.0 $M_\odot$.  The large scatter of the measurements around the hypothetical orbit could be caused by the presence of the close companion of V853 Oph A.}
\label{fig.vmotion}
\end{figure}

In Figure~\ref{fig.vmotion}, we show the motion of the primary, V853 Oph A, modeled as an unresolved source, relative to the wide tertiary, V853 Oph B.  To help guide the eye, we plot a model orbit with a period of 600 years and a total mass of 1.0~$M_{\odot}$ to approximate the motion of V853 Oph A relative to V853 Oph B over the course of the observations.  The measurements are clearly scattered around the model orbit.  This scatter could be caused by the orbital signature of the unresolved companion shifting the photocenter of V853 Oph A.  We tried fitting a sinusoidal function to the residuals but were not able to derive a significant periodicity.  With improvements in the sensitivity limits of ground-based infrared interferometers, resolving the close companion of V853 Oph A could prove to be an ideal target for interferometry.

\section{Total Mass Estimates}
\label{sect.mass}

\begin{figure}
	\begin{center}
	\scalebox{0.44}{\includegraphics{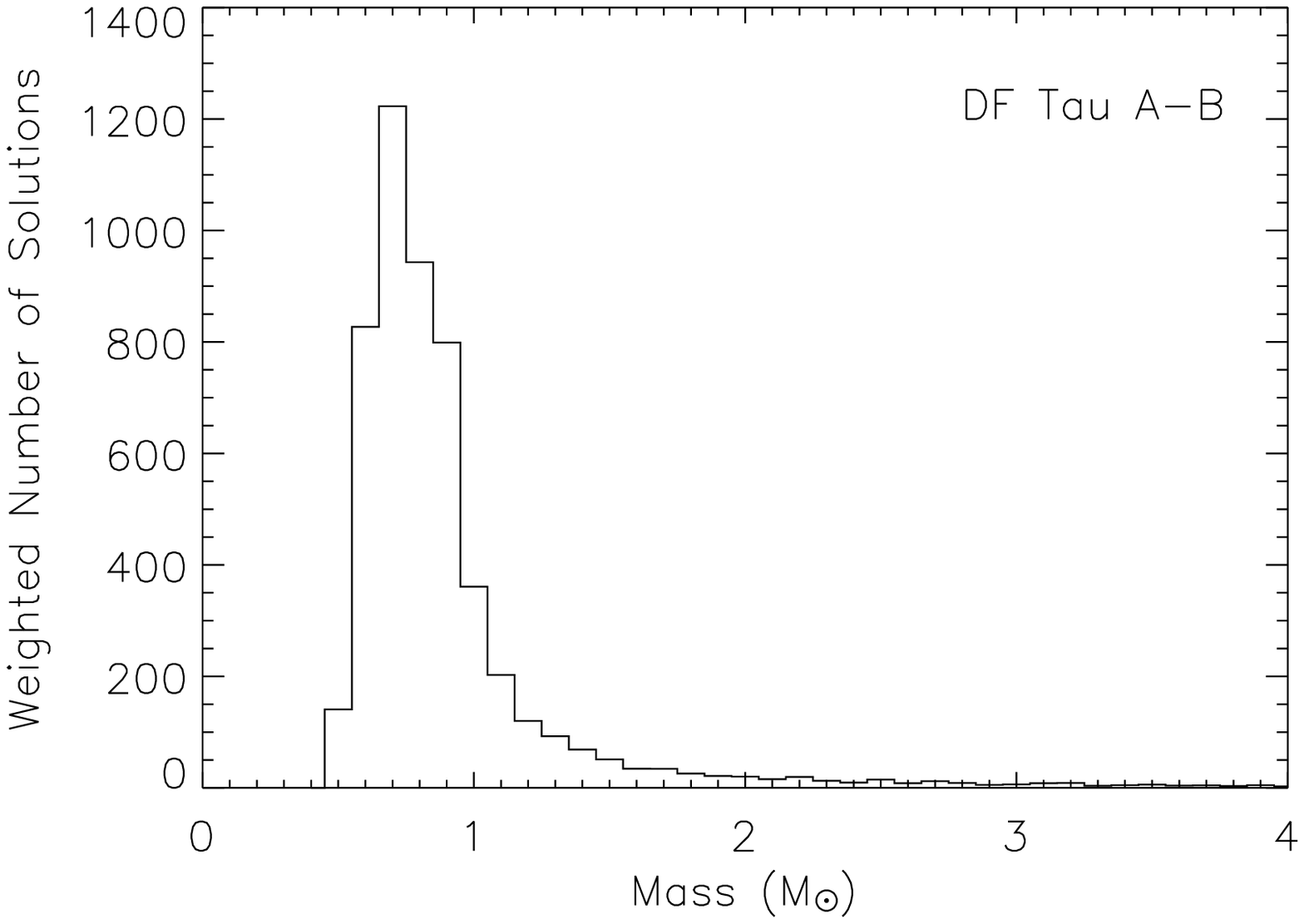}}
	\scalebox{0.44}{\includegraphics{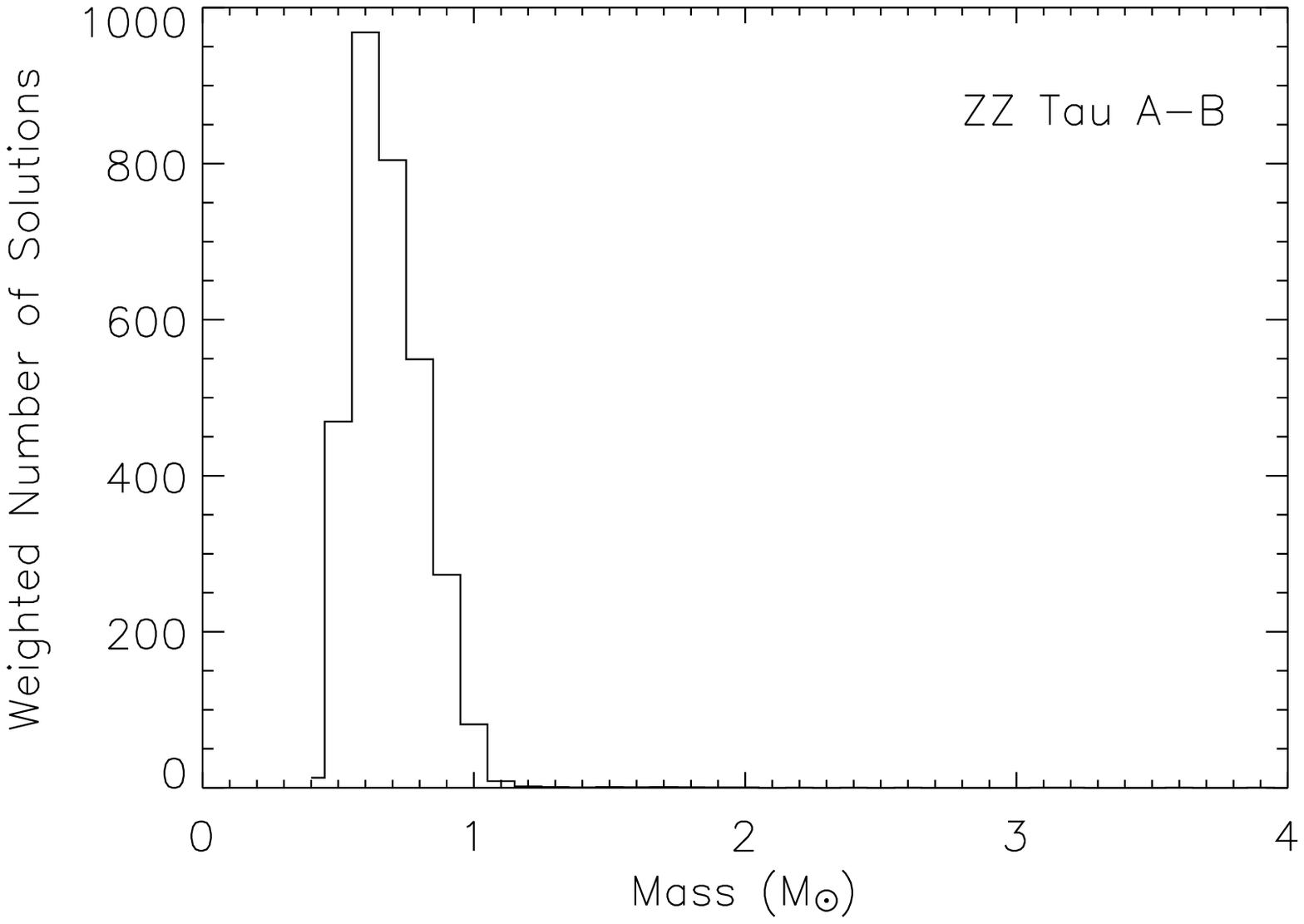}} \\
	\scalebox{0.44}{\includegraphics{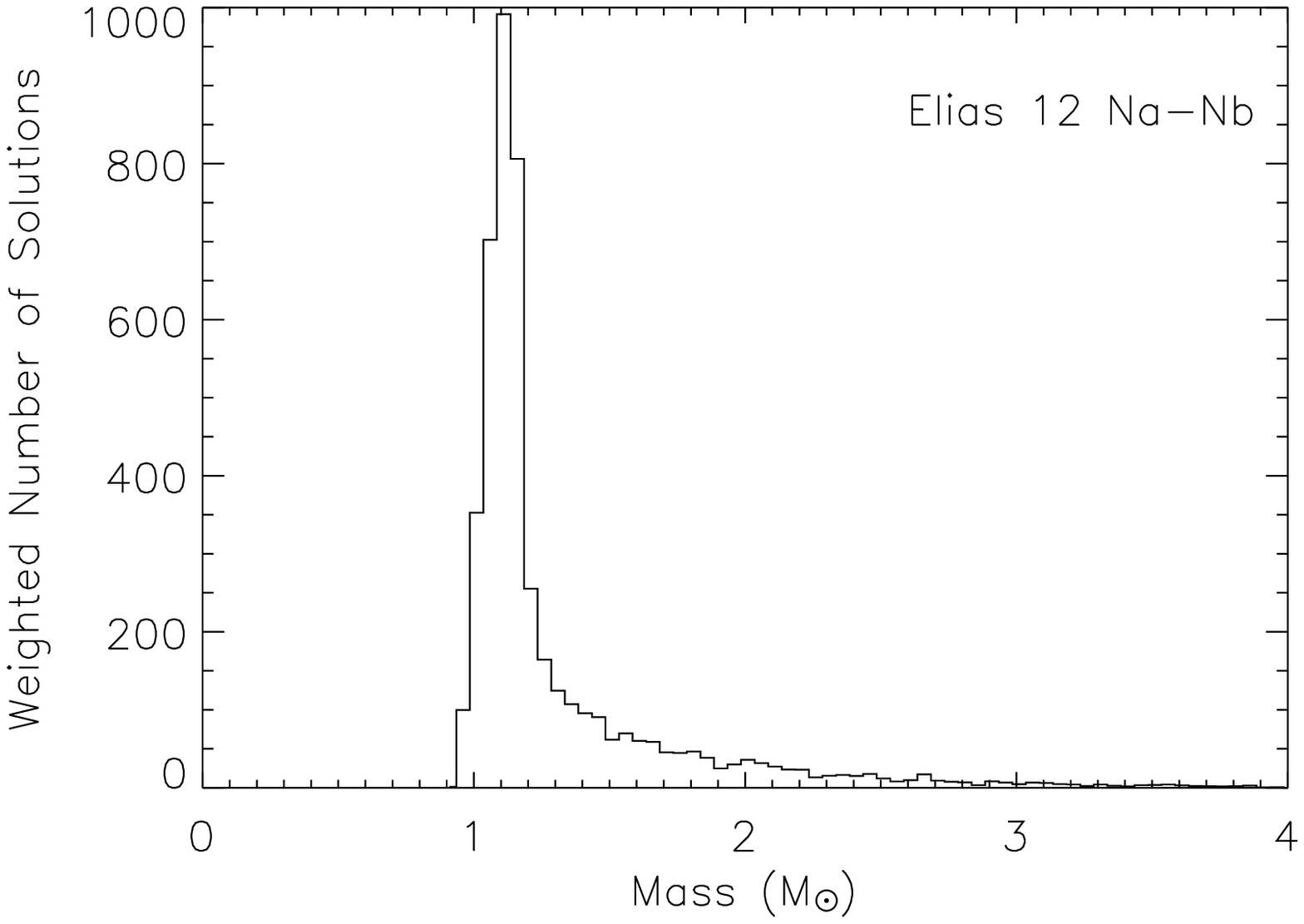}}
	\scalebox{0.44}{\includegraphics{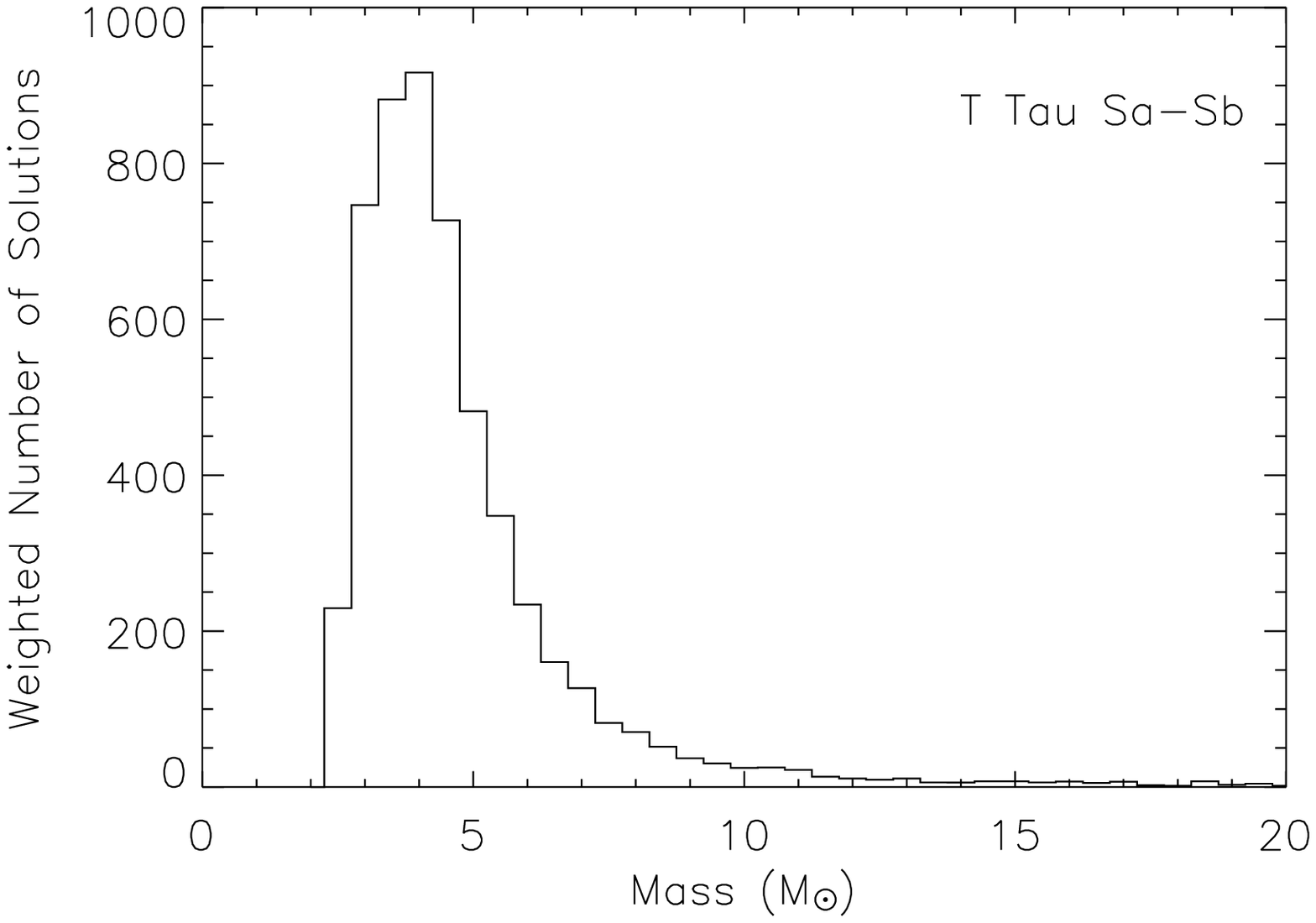}}
	\end{center}
   \caption[Mass distribution]{Weighted distributions for the total mass of DF Tau (top left), ZZ Tau (top right), Elias 12 Na-Nb (bottom left), and T Tau Sa-Sb (bottom right).  These results were obtained by searching for orbital solutions within the $3 \sigma$ confidence interval ($\Delta\chi^2 = 9$) by using our Monte Carlo grid search technique.  The total mass ($a^3/P^2$) is derived for each of the 10,000 orbits found within this interval.  In the histograms, each solution is weighted by its $\chi^2$ probability.}
\label{fig.masshisto}
\end{figure}

The total mass of a binary is determined by $a^3/P^2$, with the semi-major axis on a physical scale.  To examine the distribution of masses allowed by the current set of orbital measurements, we used our Monte Carlo grid search technique to explore the 3-$\sigma$ confidence interval ($\Delta \chi^2 = 9$) for DF Tau, ZZ Tau, Elias 12 Na-Nb, and T Tau Sa-Sb.  As in the previous section, we searched for orbits out to periods of 500 years and eccentricities of 0.99.  After finding 10,000 possible orbits that fit the data, we computed the total mass for each orbital solution, $M_{tot} = (a/1AU)^3/(1yr/P)^2 ~M_\odot$.  To convert the angular separations to physical sizes, we used the average distance to the Taurus star forming region, d=140 pc \citep{kenyon94}.  For T Tau, we used the VLBA parallax measurement of $141.5 \pm 2.8$ pc \citep{loinard05}.  The resulting mass distributions for DF Tau, ZZ Tau, Elias 12 Na-Nb, and T Tau Sa-Sb are shown in Figure~\ref{fig.masshisto}.  In these histograms, each orbital solution is weighted by its $\chi^2$ probability.  We truncated the histograms for DF Tau and T Tau at 4 and 10 $M_{\odot}$ respectively; masses beyond these cut-off values are not representative of the measured luminosities and spectral types of the systems.  This condition tends to eliminate highly eccentric, short period orbits from the sample and typically removes less than 5\% of the 10,000 solutions in the 3-$\sigma$ interval.  There is a clearly defined peak in each of the mass distributions.  The last row of Table~\ref{tab.orbpar} lists the most probable value for the total masses of DF Tau, ZZ Tau, Elias 12 Na-Nb, and T Tau Sa-Sb.  Since the mass distributions are asymmetric, the reported values represent the median of the distributions.  The range of the quoted uncertanties include 34\% of the values on each side of the median.  For ZZ Tau, where we computed a preliminary orbit solution, the dynamical mass estimate based on the weighted Monte Carlo distribution agrees with the value derived from the formal fit within 1$\sigma$.  For Elias 12 Na-Nb, where we found two possible orbit solutions based on the combined AO, FGS, and LO data, the dynamical mass determined from the weighted Monte Carlo distribution favors the value derived from the 11.85~year orbit.

\begin{figure}
   \scalebox{1.1}{\includegraphics{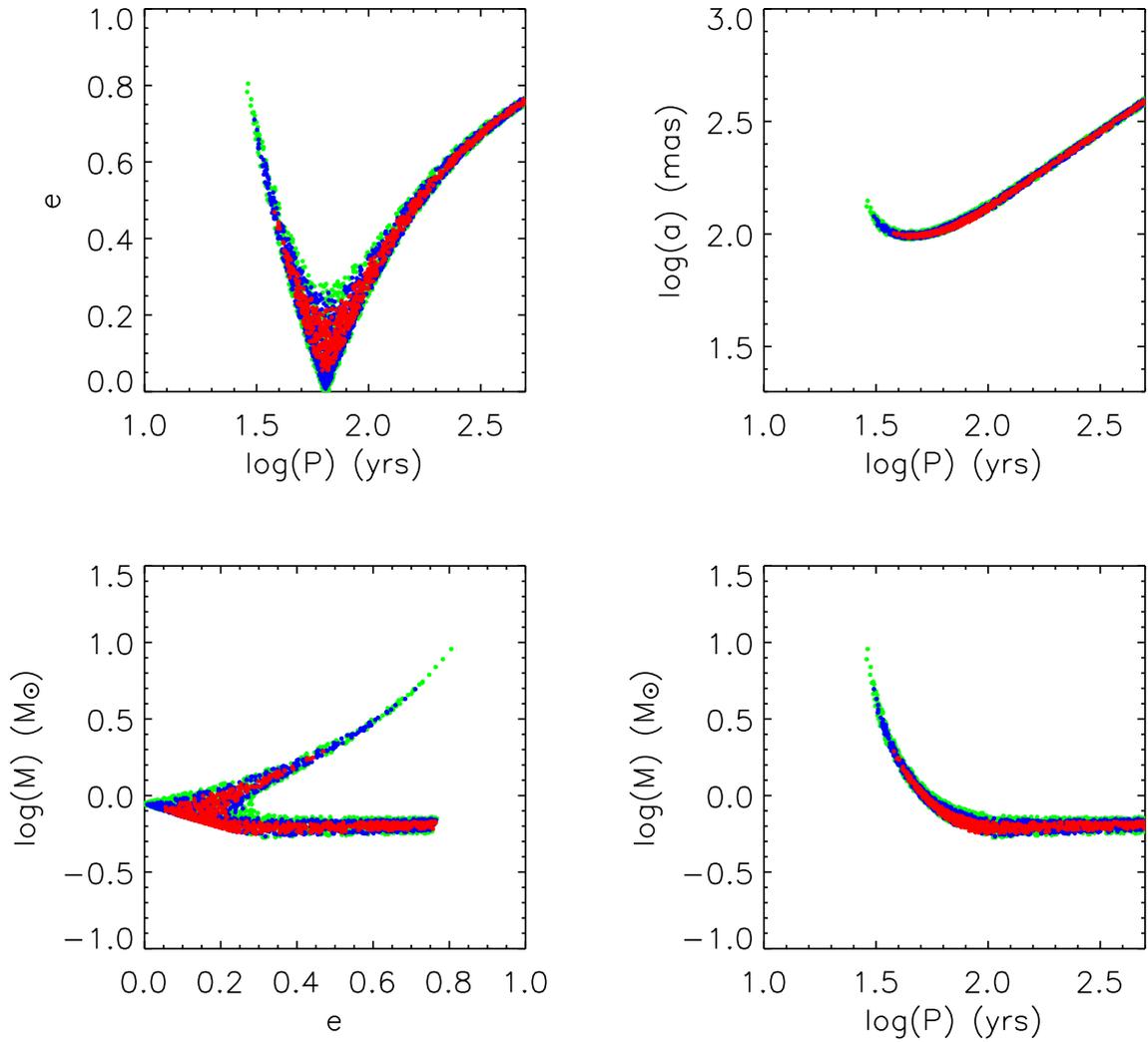}}
   \caption[DF Tau $\chi^2$ surface]{Crosscuts through the $\chi^2$ surface for DF Tau.  The color codes correspond to the $1 \sigma$ (red), $2\sigma$ (blue), and $3 \sigma$ (green) confidence intervals.}
\label{fig.dfchi}
\end{figure}

To further examine the typical shape of the allowed parameter space, Figure~\ref{fig.dfchi} shows crosscuts through the $\chi^2$ surface for different sets of orbital parameters for DF Tau.  The plot of angular semi-major axis versus the orbital period shows the strong correlation between $a$ and $P$ that produces the well-defined distributions for the total mass.  The plots of the total mass versus $e$ and $P$, as well as the plot of $P$ versus $e$, show the long tail of short period, highly eccentric solutions that give rise to unphysically large values of the total mass.  It is interesting to note that even if the chosen search ranges are extended beyond $P$=500 years, the calculated value for the total mass remains unchanged since the mass distribution flattens out at large periods, converging to a value corresponding to the computed median mass.

The result that poorly determined orbits can produce reliable values for the total mass was noted by \citet{eggen67} and \citet{hartkopf01}.  The system mass is measured from the instantaneous radial acceleration and separation of the stars.  Although the complete orbit may not be known, the orbital solutions must reproduce the projected radial acceleration of the system at the current separation, measured through the curvature of the apparent orbit.  As a result, the value of $(a^3/P^2)$ can be determined with much greater accuracy than the independent uncertainties in the orbital elements suggest.

\section{Discussion}

\subsection{Comments on Component Masses}

\begin{figure}
   \scalebox{1.0}{\includegraphics{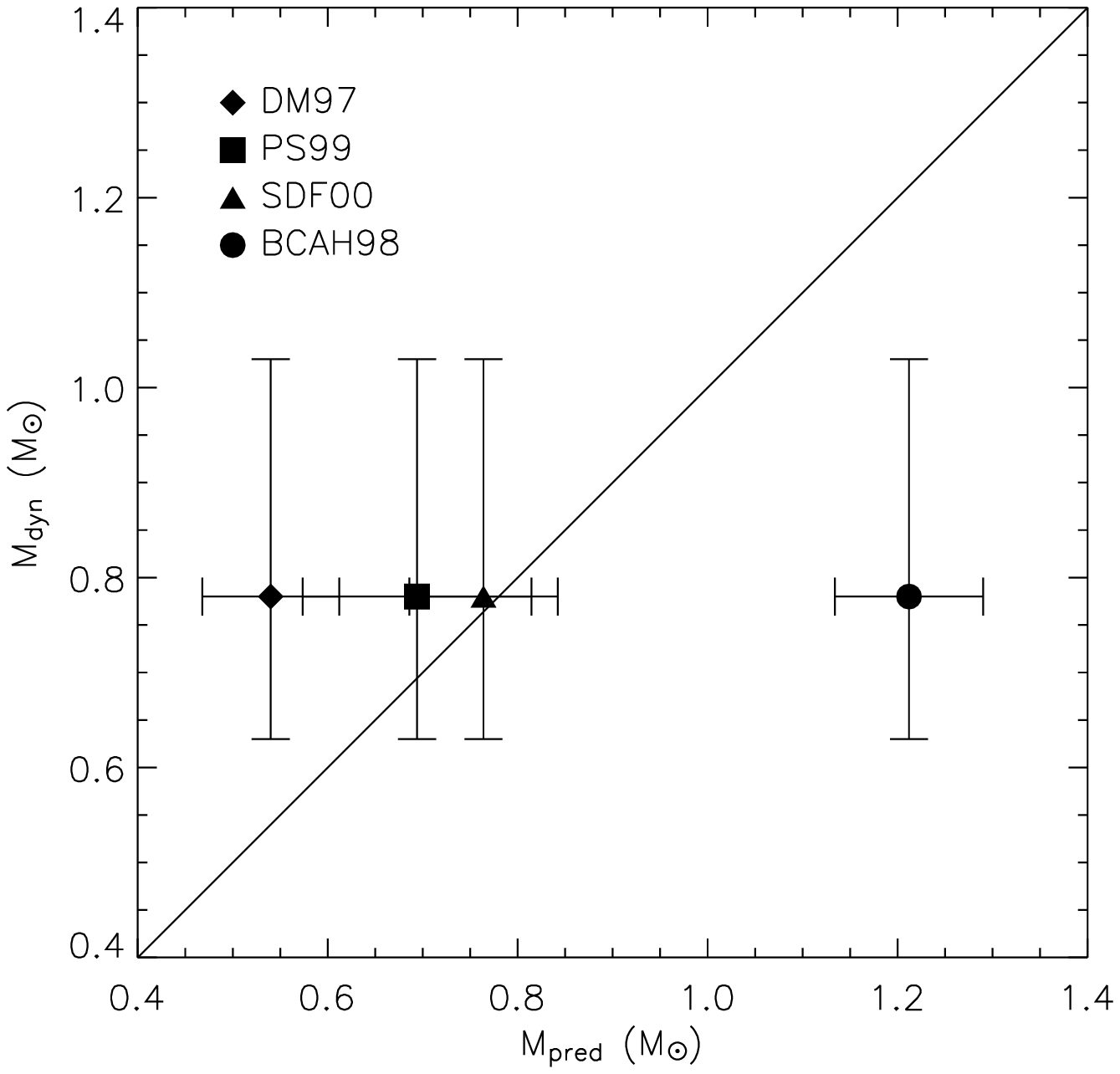}}
   \caption{Total dynamical mass of the DF Tau binary compared with masses predicted from the evolutionary tracks of  \citet[DM97]{dantona97}, \citet[BCAH98]{baraffe98}, \citet[PS99]{palla99}, and \citet[SDF00]{siess00}.  The solid line shows where the predicted and dynamical mass estimates would be equal.}
\label{fig.compmass}
\end{figure}

DF Tau is the only binary in our sample for which there is a spectral type and luminosity measured for both the primary and secondary components.  Using the evolutionary tracks computed by \citet[DM97]{dantona97}, \citet[BCAH98]{baraffe98}, \citet[PS99]{palla99}, and \citet[SDF00]{siess00} and the luminosities and spectral types listed in Table~\ref{tab.targlist}, we compute the predicted total mass of the DF Tau binary for each set of tracks.  To convert the spectral types to effective temperatures, we use the temperature scale conversion of \citet{kenyon95}.  The age of the DF Tau binary predicted from the different sets of evolutionary tracks ranges from 0.5 to 2 Myr.  In Figure~\ref{fig.compmass}, we compare the masses predicted from the theoretical tracks with our dynamical mass estimate of 0.78$^{+0.25}_{-0.15} M_\odot $ ~measured for DF Tau.  The error bars for the predicted masses are derived from an uncertainty of $\pm1$~subclass in the spectral type determination.  The predicted masses from each of the tracks agree with our dynamical estimate within $\sim 1.5\sigma$.  Based on the current uncertainties, it is premature to make a qualitative assessment of the evolutionary tracks based only on the total mass estimate of DF Tau.  We are working on a program to determine the spectral types and luminosities of the close companions in ZZ Tau and Elias 12 N.

For an age of 1~Myr, the M1 spectral type of T Tau Sb \citep{duchene02} corresponds to a predicted mass of 0.4-0.7~$M_\odot$, depending on the choice of theoretical tracks.  Combined with the total dynamical mass measurement of 4.13$^{+1.58}_{-0.97} M_\odot$ for the T Tau S binary, this range implies a predicted mass of 2.4-5.3~$M_\odot$ for T Tau Sa.  Thus, T Tau Sa may be a Herbig AeBe star caught at a very early phase of its evolution.  With an estimated mass of 2.1~$M_{\odot}$ for T Tau N \citep{white01}, our dynamical results suggest that T Tau Sa is the most massive component in the system.  This is consistent with the photometric and spectroscopic findings in the literature \citep{koresko97,beck04,duchene05} and the orbit fits based on radio and infrared data \citep{johnston04, duchene06}.

\subsection{Variability}

\begin{figure}
   \scalebox{0.78}{\includegraphics{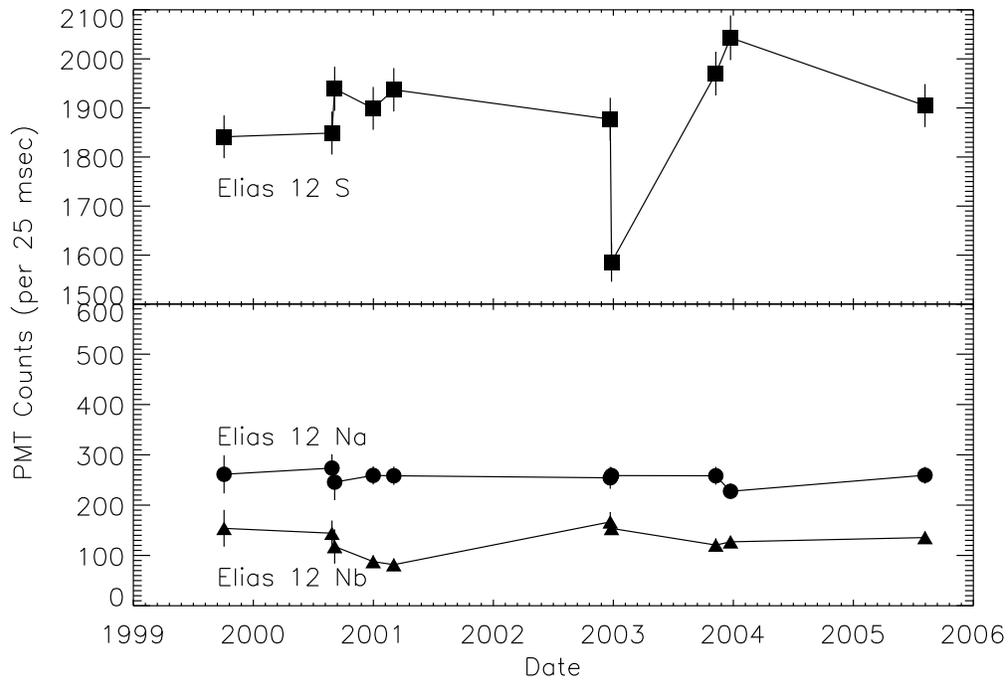}}
   \caption[Photometric variability of Elias 12]{Variability of the FGS PMT counts measured for the three components in Elias 12.}
\label{fig.el12mag}
\end{figure}

\begin{figure}
   \scalebox{0.78}{\includegraphics{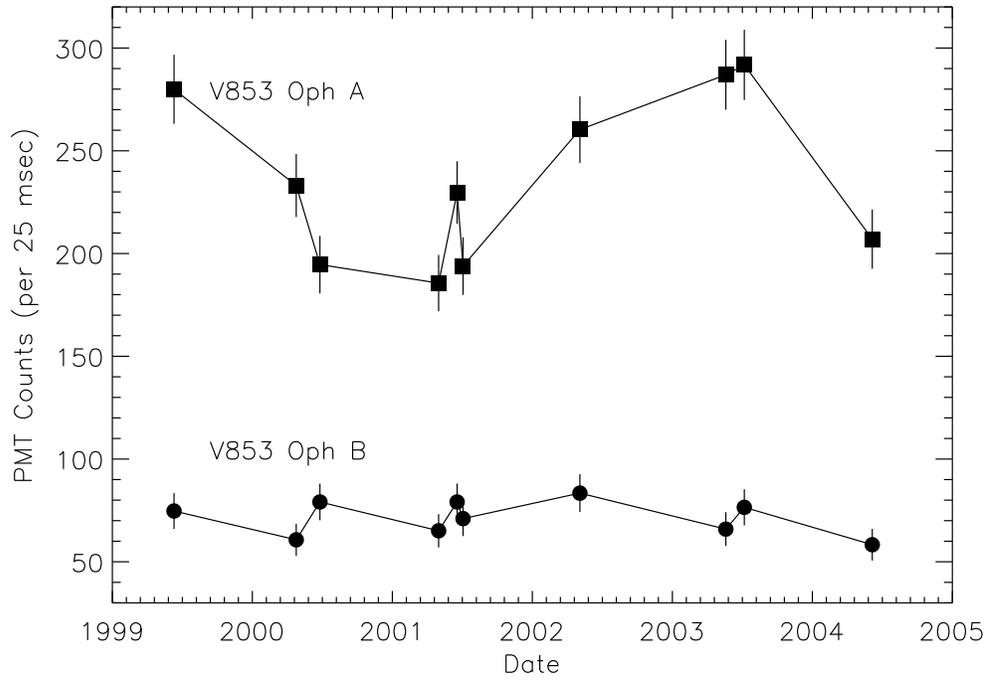}}
   \caption[Photometric variability of V853 Oph]{Variability of the FGS PMT counts measured for V853 Oph A and B.}
\label{fig.v853ophmag}
\end{figure}

The two recent FGS flux measurements for DF Tau presented in Table~\ref{tab.fgscounts} are consistent with the results in \citet{schaefer03}, where we found that the flux of the DF Tau primary varies while the secondary remains constant.  In Figures~\ref{fig.el12mag} and~\ref{fig.v853ophmag}, we plot the changes in the FGS counts measured for the components in the Elias 12 triple and for V853 Oph A-B.  As with DF Tau, it appears that the primaries, Elias 12 S and V853 Oph A, show the most variability in the system.  It is interesting that in all three systems, the primaries display the activity associated with accretion from a disk.

\begin{figure}
   \scalebox{1.0}{\includegraphics{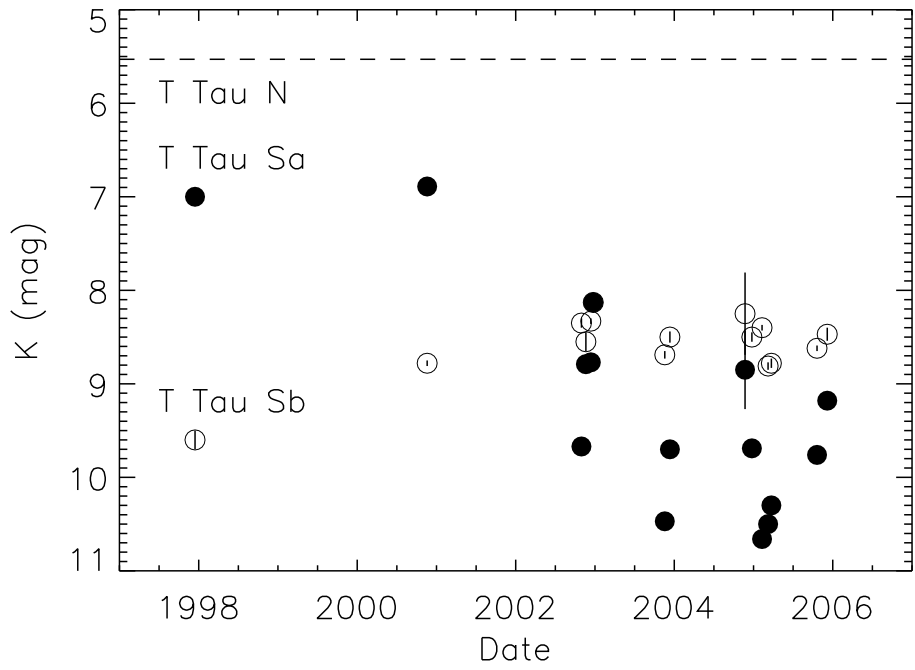}}
   \caption[Photometric variability of T Tau]{Cumulative K-band variability measured for T Tau Sa (closed circles) and T Tau Sb (open circles).  We assume a constant magnitude of K=5.53 for T Tau N (dashed line; Beck et al. 2004).}
\label{fig.ttmag}
\end{figure}

In Figure~\ref{fig.ttmag}, we show the cumulative magnitude variations observed for the components of T Tau, based on our measurements presented in Table~\ref{tab.ttaumag} and measurements reported in the literature \citep{koresko00,kohler00,duchene02,furlan03,beck04,duchene05,mayama06}.  These results show that the large variability of T Tau Sa continues in the recent measurements, with variations of $\sim$ 2 mag taking place over the course of a few months.  \citet{duchene05} speculate that this variability may be caused by a clumpy, nearly edge-on circumstellar disk rotating around the central star.

\section{Summary}

1. We report recent measurements of the orbital motion in DF Tau, ZZ Tau, Elias 12, T Tau, and V853 Oph using the Fine Guidance Sensors on {\it HST} and adaptive optics imaging at the Keck and Gemini Observatories.  

2. Based on the cumulative set of orbital measurements, we find a likely orbital solution with a period of $\sim$ 11.9 years for the Elias 12 Na-Nb binary.  Similarly, by combining our FGS and AO measurements of ZZ Tau with the published lunar occultation observation, we find a preferred orbital solution with a period of $\sim$ 32 years.  Our analyses for DF Tau and T Tau Sa-Sb show that we are not yet able to determine definitive orbital parameters for these systems.

3. We developed a technique to determine dynamical mass estimates for binary orbits with limited orbital coverage.  By using a Monte Carlo technique to search the orbital parameter space, we constructed weighted distributions for the total mass in DF Tau, ZZ Tau, Elias 12 Na-Nb, and T Tau Sa-Sb.  We find total masses of 0.78$^{+0.25}_{-0.15} M_\odot$, 0.66$^{+0.15}_{-0.11} M_\odot$, 1.13$^{+0.36}_{-0.09} M_\odot$, and 4.13$^{+1.58}_{-0.97} M_\odot$, respectively, for these systems by adopting a distance of 140~pc.

4. The uncertainties in our dynamical mass estimates, combined with the limited knowledge of the luminosites and effective temperatures of the close companions preclude us from making a qualitative assessment of the theoretical tracks of PMS evolution based only on the masses presented herein.  With continued orbit mapping and precise distance measurements obtained through astrometric missions such as SIM and GAIA, we expect to measure the total dynamical masses of the binaries to better than 5\%.  Determining the luminosity and effective temperature of the the close companions will be the next step toward the goal of testing the evolutionary tracks.

\acknowledgements

We thank D. Peterson for many helpful discussions on the statistical analysis developed in this paper.  We also thank G. Sprouse for valuable suggestions on analyzing the mass distributions, B. Oppenheimer for providing constructive feedback on the thesis of G.H.S. which formed the basis for this paper, W. I. Hartkopf for generously sharing his binary orbit code, and our referee, V. Makarov, for reading the manuscript and providing helpful comments.  This work was supported by NSF grant AST 02-05427 (M.S. and G.H.S) and JPL contract 1276601 (G.H.S.).  This research is based, in part, on observations made with the NASA/ESA Hubble Space Telescope, obtained at the Space Telescope Science Institute, which is operated by the Association of Universities for Research in Astronomy, Inc., under NASA contract NAS 5-26555.  The {\it HST} observations are associated with GO Proposals 6486, 7487, 8339, 8616, 8783, 9229, and 9335.  We also presented data obtained at the W.M. Keck Observatory, which is operated as a scientific partnership among the California Institute of Technology, the University of California, and the National Aeronautics and Space Administration. The Observatory was made possible by the generous financial support of the W.M. Keck Foundation.  We also based this research on observations obtained at the Gemini Observatory, which is operated by the Association of Universities for Research in Astronomy, Inc., under a cooperative agreement with the NSF on behalf of the Gemini partnership: the National Science Foundation (United States), the Particle Physics and Astronomy Research Council (United Kingdom), the National Research Council (Canada), CONICYT (Chile), the Australian Research Council (Australia), CNPq (Brazil) and CONICET (Argentina).  The Gemini observations are associated with the programs GN-2004A-Q-27-9 and GN-2004B-Q-7.  This research has made use of the Washington Double Star Catalog maintained at the U.S. Naval Observatory and the SIMBAD database operated at CDS, Strasbourg, France.  We wish to recognize and extend our gratitude to the Hawaiian community for the opportunity to conduct these observations from the summit of Mauna Kea.

\clearpage

\begin{deluxetable}{lrcclllll} 
\tablewidth{0pt}
\tablecaption{Summary of Target Properties} 
\tablehead{\colhead{Object} & \colhead{HBC} & \colhead{Coordinates (J2000)} & \colhead{V} & \colhead{B-V} & \colhead{K} & \colhead{SpT} & \colhead{L (L$_{\odot}$)} & \colhead{Ref}}
\startdata 
DF Tau A   & 36  & 04:27:02.80  +25:42:22.9 & 12.43\tablenotemark{v} & 0.75 & 7.0 & M2.0 & 0.56 & 1,1,2,3,4 \\
DF Tau B   &     &                          & 13.10 & 1.37 & 7.9 & M2.5 & 0.69 & 1,1,2,3,4 \\
ZZ Tau\tablenotemark{\dag}     & 46  & 04:30:51.40  +24:42:22.2 & 14.34 & 1.52 & 8.5 & M3   & 0.7\tablenotemark{*} & 5,5,6,5,7 \\
Elias 12 S & 404 & 04:33:06.62  +24:09:55.2 & 11.56 & 1.18 & 7.2 & K5   & 2.3 & 1,1,6,3,3 \\
Elias 12 N\tablenotemark{\dag} &     &                          & 13.35 & 1.24 & 8.3 & M2.0 & 1.04\tablenotemark{*} & 1,1,6,3,3 \\
T Tau N    & 35  & 04:21:59.42  +19:32:06.4 &  9.90 & 1.19 & 5.53 & K0  & 12.5 & 5,5,1,8,7 \\
T Tau Sa   &     &                          &       &      & 9.7\tablenotemark{v}  &      &  & 9 \\
T Tau Sb   &     &                          &       &      & 8.5\tablenotemark{v}  & M1   &  & 9,10 \\
V853 Oph\tablenotemark{\ddag}   & 266 & 16:28:45.29  -24:28:18.4 & 13.3  & 1.1  & 8.0 & M2.5 &  & 11,11,6,11  \\
\
\enddata
\tablenotetext{\dag}{Binary}
\tablenotetext{\ddag}{Triple}
\tablenotetext{v}{Variable}
\tablenotetext{*}{Total luminosity of binary}
\tablerefs{(1) White \& Ghez 2001; (2) Chen et al. 1990; (3) Hartigan \& Kenyon 2003; (4) Hillenbrand \& White 2004; (5) Herbig \& Bell 1988; (6) Simon et al. 1995; (7) Strom et al. 1989; (8) Basri \& Batalha 1990; (9) This work; (10) Duch\'{e}ne, Ghez, \& McCabe 2002; (11) Bouvier \& Appenzeller 1992}
\label{tab.targlist} 
\end{deluxetable} 

\clearpage

\begin{deluxetable}{ll}
\tablewidth{0pt}
\tablecaption{Log of Fine Guidance Sensor Observations}
\tablehead{\colhead{Object} & \colhead{UT Date}}
\startdata
DF Tau   & 2003 Jan 23, 2003 Nov 19   \\ 
ZZ Tau   & 2003 Sep 21, 2005 Nov 06   \\ 
Elias 12 & 1999 Oct 03, 2000 Aug 27, 2000 Sep 04 \\
         & 2000 Dec 31, 2001 Mar 04, 2002 Dec 22 \\
         & 2002 Dec 26, 2003 Nov 08, 2003 Dec 23 \\
         & 2005 Aug 06  \\  
V853 Oph & 1999 Jun 10, 2000 Apr 24, 2000 Jun 25 \\ 
         & 2001 May 01, 2001 Jun 18, 2001 Jul 04 \\
         & 2002 May 04, 2003 May 20, 2003 Jul 07 \\
         & 2004 Jun 05 \\
\enddata
\label{tab.fgs-obs}
\end{deluxetable}

\clearpage

\begin{deluxetable}{llllcccc}
\tablewidth{0pt}
\tablecaption{Log of Adaptive Optics Observations}
\tablehead{\colhead{Object} & \colhead{Date} & \colhead{Detector} & \colhead{Filt} & \colhead{AO Rate} & \colhead{T$_{int}$} & \colhead{No.} & \colhead{No.} \\ \colhead{} & \colhead{} & \colhead{} & \colhead{} & \colhead{(Hz)} & \colhead{(s)} & \colhead{Co-add} & \colhead{Im.}}
\startdata
Elias 12 & 2002 Mar 8   & NIRC2 & H    & 272 & 0.25 & 10 & 5 \\
         &              &       & K$'$ & 272 & 0.25 & 10 & 10 \\
         & 2002 Oct 30  & NIRC2 & H    & 140 & 0.20 & 20 & 10 \\
         &              &       & H    & 120 & 0.18 & 10 & 5 \\
         &              &       & K$'$ & 140 & 0.50 & 10 & 10 \\
         &              &       & K$'$ & 120 & 0.18 & 10 & 5 \\
         & 2004 Dec 24  & NIRC2 & H    & 420 & 0.18 & 10 & 10 \\
         &              &       & K$'$ & 420 & 0.18 & 10 & 10 \\
         & 2005 Mar 1   & NIRI  & Hcont & 200 & 1.00 & 20 & 25 \\
         &              &       & Kcont & 200 & 1.00 & 20 & 25 \\
         & 2005 Dec 6   & NIRC2 & H     & 272 & 0.18 & 10 &  5 \\
         &              &       & K$'$  & 272 & 0.18 & 10 &  5 \\
\noalign{\vskip 1.0ex}%
\hline
\noalign{\vskip 1.0ex}%
T Tau    & 2002 Oct 30  & NIRC2  & K$'$  &  154 & 0.18 & 10 & 15 \\
         &              &        & Br$\gamma$  &  332 & 0.30 & 10 & 15 \\
         & 2003 Nov 18  & NIRI   & Kcont & 1000 & 0.25 & 10 & 10 \\ 
         & 2004 Dec 24  & NIRC2  & Kcont &  660 & 0.18 & 10 & 20 \\
         & 2005 Feb 9   & NIRI   & Kcont & 1000 & 0.25 & 10 &  5 \\
         & 2005 Mar 9   & NIRI   & Kcont & 1000 & 0.20 & 10 &  5 \\
         & 2005 Mar 24  & NIRI   & Kcont & 1000 & 0.20 & 90 & 30 \\
         &              &        & H2(1-0) & 1000 & 0.20 & 90 & 35 \\
         & 2005 Oct 20  & NIRI   & Kcont & 1000 & 0.20 & 20 & 10 \\
         & 2005 Dec 6   & NIRC2  & K$'$  & 90   & 0.18 & 10 & 15 \\
\noalign{\vskip 1.0ex}%
\hline
\noalign{\vskip 1.0ex}%
V853 Oph & 2002 Mar 8  & NIRC2  & H    & 79 & 0.5 & 10 & 10 \\
         &             &        & K$'$ & 79 & 0.5 & 10 & 10 \\
         & 2003 Apr 15 & NIRC2  & H    & 74 & 1.0 & 10 & 20 \\
         &             &        & K$'$ & 74 & 1.0 & 10 & 20 \\
         & 2003 Apr 20 & SCAM   & K$'$ & 91 & 0.5 & 5 & 8 \\
         & 2004 Apr 3  & NIRI   & FeII & 1000  & 1.2 & 10 & 20 \\
         &             &        & Br$\gamma$ & 1000 & 1.0 & 10 & 20 \\
         & 2004 Apr 8  & NIRI   & FeII & 1000  & 1.2 & 10 & 20 \\
         &             &        & Br$\gamma$ & 1000 & 1.0 & 10 & 20 \\
         & 2004 Jun 15 & NIRC2  & H    & 55 & 1.0 & 10 & 10  \\
         &             &        & K$'$ & 55 & 1.0 & 10 & 10  \\
         & 2004 Jul 14 & NIRC2  & H    & 55 & 0.18 & 10 & 15 \\
         &             &        & K$'$ & 55 & 0.18 & 10 & 10 \\
\noalign{\vskip 1.0ex}%
\hline
\noalign{\vskip 1.0ex}%
DF Tau   & 2004 Dec 24  & NIRC2 & Kcont & 235 & 0.18 & 10 & 10 \\

         & 2005 Dec 6   & NIRC2 & H     & 161 & 0.40 & 10 &  5 \\
         &              & NIRC2 & K$'$  & 161 & 0.40 & 10 & 15 \\
         &              & NIRC2 & K$'$  &  84 & 0.40 & 10 &  5 \\
\noalign{\vskip 1.0ex}%
\hline
\noalign{\vskip 1.0ex}%
ZZ Tau   & 2004 Dec 24  & NIRC2 & K$'$ &  73 & 0.40 & 10 & 10 \\
\enddata
\label{tab.ao-obs}
\end{deluxetable}

\clearpage

\begin{deluxetable}{llll} 
\tabletypesize{\small}
\tablewidth{0pt}
\tablecaption{Recent Measurements of PMS Stars in Multiple Systems} 
\tablehead{
\colhead{Year} & \colhead{$\rho$(mas)} & \colhead{P.A.($\degr$)} & 
\colhead{Instrument}}
\startdata 
\multicolumn{4}{c}{DF Tau} \\
 \noalign{\vskip .8ex}%
 \hline
 \noalign{\vskip .8ex}%
2003.061 & 101.5$\pm$2.0 & 258.6$\pm$1.2 & FGS   \\   
2003.883 & 105.1$\pm$2.0 & 249.6$\pm$1.1 & FGS   \\   
2004.981 & 109.8$\pm$2.0 & 247.0$\pm$1.1 & NIRC2 \\  
2005.930 & 110.9$\pm$1.6 & 241.7$\pm$0.8 & NIRC2 \\  
\cutinhead{ZZ Tau}
2003.723 & 62.8$\pm$1.5 &  87.6$\pm$1.4  & FGS   \\  
2004.980 & 61.2$\pm$1.4 &  74.9$\pm$1.3  & NIRC2 \\  
2005.849 & 61.7$\pm$1.5 &  67.2$\pm$1.4  & FGS   \\  
\cutinhead{Elias 12 S-Na}
1999.756 & 282.9$\pm$2.1 & 324.31$\pm$0.43 & FGS   \\  
2000.655 & 286.9$\pm$1.8 & 323.56$\pm$0.36 & FGS   \\  
2000.677 & 284.9$\pm$2.7 & 323.60$\pm$0.54 & FGS   \\  
2000.999 & 287.0$\pm$1.8 & 322.83$\pm$0.36 & FGS   \\  
2001.170 & 286.2$\pm$0.9 & 322.12$\pm$0.18 & FGS   \\  
2002.182 & 285.4$\pm$1.2 & 321.34$\pm$0.29 & NIRC2 \\  
2002.829 & 282.0$\pm$0.9 & 319.86$\pm$0.25 & NIRC2 \\  
2002.974 & 281.4$\pm$1.5 & 319.25$\pm$0.31 & FGS   \\  
2002.985 & 281.8$\pm$1.1 & 319.60$\pm$0.22 & FGS   \\  
2003.854 & 271.6$\pm$1.0 & 317.33$\pm$0.21 & FGS   \\  
2003.977 & 273.2$\pm$1.0 & 316.96$\pm$0.21 & FGS   \\  
2005.163 & 268.9$\pm$1.0 & 312.69$\pm$0.22 & NIRI  \\  
2005.597 & 260.5$\pm$0.7 & 310.92$\pm$0.15 & FGS   \\  
2005.930 & 259.2$\pm$1.0 & 310.31$\pm$0.27 & NIRC2 \\  
\cutinhead{Elias 12 S-Nb}
1999.756 & 305.0$\pm$4.4 & 321.29$\pm$0.83 & FGS   \\  
2000.655 & 284.6$\pm$4.1 & 316.81$\pm$0.83 & FGS   \\  
2000.677 & 285.2$\pm$4.6 & 317.56$\pm$0.92 & FGS   \\  
2000.999 & 282.9$\pm$4.2 & 315.60$\pm$0.85 & FGS   \\  
2001.170 & 279.4$\pm$3.6 & 314.66$\pm$0.74 & FGS   \\  
2002.182 & 259.5$\pm$1.4 & 314.79$\pm$0.34 & NIRC2 \\  
2002.829 & 248.5$\pm$0.9 & 313.52$\pm$0.26 & NIRC2 \\  
2002.974 & 247.0$\pm$2.2 & 312.80$\pm$0.51 & FGS   \\  
2002.985 & 248.9$\pm$2.4 & 313.14$\pm$0.55 & FGS   \\  
2003.854 & 227.5$\pm$2.4 & 312.98$\pm$0.60 & FGS   \\  
2003.977 & 228.8$\pm$2.4 & 313.58$\pm$0.60 & FGS   \\  
2005.163 & 220.8$\pm$1.3 & 311.63$\pm$0.33 & NIRI  \\  
2005.597 & 214.8$\pm$1.3 & 311.32$\pm$0.35 & FGS   \\  
2005.930 & 214.8$\pm$1.0 & 311.56$\pm$0.27 & NIRC2 \\  
\cutinhead{Elias 12 Na-Nb}
1999.756 & 27.0$\pm$4.9 & 287.8$\pm$10.4 & FGS   \\  
2000.655 & 33.7$\pm$4.5 & 226.3$\pm$7.6  & FGS   \\  
2000.677 & 30.0$\pm$5.3 & 231.2$\pm$10.2 & FGS   \\  
2000.999 & 36.2$\pm$4.6 & 222.7$\pm$7.2  & FGS   \\  
2001.170 & 37.4$\pm$3.7 & 217.8$\pm$5.7  & FGS   \\  
2002.182 & 40.4$\pm$1.7 & 188.4$\pm$2.4  & NIRC2 \\  
2002.829 & 44.5$\pm$1.1 & 177.9$\pm$1.3  & NIRC2 \\  
2002.974 & 45.4$\pm$2.7 & 177.0$\pm$3.4  & FGS   \\  
2002.985 & 44.5$\pm$2.6 & 178.6$\pm$3.4  & FGS   \\  
2003.854 & 47.9$\pm$2.6 & 158.4$\pm$3.1  & FGS   \\  
2003.977 & 46.8$\pm$2.6 & 153.8$\pm$3.2  & FGS   \\  
2004.979 & 47.7$\pm$1.9 & 139.7$\pm$2.3  & NIRC2 \\  
2005.163 & 48.3$\pm$1.6 & 137.6$\pm$1.9  & NIRI  \\  
2005.597 & 45.8$\pm$1.5 & 129.0$\pm$1.9  & FGS   \\  
2005.930 & 44.7$\pm$1.3 & 124.3$\pm$1.5  & NIRC2 \\  
\tablebreak
\cutinhead{T Tau N-Sa}
2002.829 & 697.6$\pm$2.1 & 183.25$\pm$0.26 & NIRC2 \\  
2003.881 & 700.4$\pm$2.9 & 183.15$\pm$0.55 & NIRI  \\  
2004.979 & 698.5$\pm$2.2 & 184.74$\pm$0.27 & NIRC2 \\  
2005.108 & 704.0$\pm$1.6 & 184.42$\pm$0.14 & NIRI  \\  
2005.184 & 702.8$\pm$2.4 & 184.36$\pm$0.20 & NIRI  \\  
2005.225 & 703.7$\pm$0.9 & 184.27$\pm$0.09 & NIRI  \\  
2005.801 & 703.3$\pm$1.2 & 274.91$\pm$0.11 & NIRI  \\  
2005.930 & 703.0$\pm$5.7 & 185.79$\pm$0.51 & NIRC2 \\  
\cutinhead{T Tau N-Sb}
2002.829 & 684.8$\pm$1.6   & 192.02$\pm$0.24 & NIRC2 \\  
2003.881 & 673.87$\pm$0.89 & 192.62$\pm$0.51 & NIRI  \\  
2004.979 & 662.13$\pm$1.9  & 194.34$\pm$0.25 & NIRC2 \\  
2005.108 & 660.20$\pm$0.92 & 193.89$\pm$0.09 & NIRI  \\  
2005.184 & 658.47$\pm$1.7  & 193.79$\pm$0.16 & NIRI  \\  
2005.225 & 659.26$\pm$0.78 & 193.80$\pm$0.09 & NIRI  \\  
2005.801 & 653.74$\pm$0.36 & 284.50$\pm$0.06 & NIRI  \\  
2005.930 & 652.65$\pm$3.7  & 195.19$\pm$0.38 & NIRC2 \\  
\cutinhead{T Tau Sa-Sb}
2002.829 & 106.5$\pm$2.6  & 284.5$\pm$1.4   & NIRC2 \\  
2003.881 & 116.5$\pm$3.1  & 291.0$\pm$1.6   & NIRI  \\  
2004.979 & 119.6$\pm$2.8  & 297.2$\pm$1.4   & NIRC2 \\  
2005.108 & 120.7$\pm$1.9  & 300.36$\pm$0.88 & NIRI  \\  
2005.184 & 120.3$\pm$2.9  & 300.63$\pm$1.40 & NIRI  \\  
2005.225 & 121.5$\pm$1.2  & 300.42$\pm$0.58 & NIRI  \\  
2005.801 & 123.7$\pm$1.3  & 303.25$\pm$0.58 & NIRI  \\  
2005.930 & 121.9$\pm$6.8  & 304.8$\pm$3.2   & NIRC2 \\  
\cutinhead{V853 Oph A-B}
1999.440 & 365.98$\pm$1.30 & 95.53$\pm$0.20 & FGS   \\ 
2000.313 & 359.12$\pm$0.50 & 95.01$\pm$0.08 & FGS   \\ 
2000.483 & 358.35$\pm$0.50 & 94.80$\pm$0.08 & FGS   \\ 
2001.331 & 353.33$\pm$0.63 & 94.71$\pm$0.10 & FGS   \\ 
2001.462 & 351.73$\pm$0.63 & 95.22$\pm$0.10 & FGS   \\ 
2001.505 & 351.95$\pm$0.50 & 94.90$\pm$0.08 & FGS   \\ 
2002.340 & 347.11$\pm$0.67 & 94.69$\pm$0.11 & FGS   \\ 
2003.286 & 344.68$\pm$2.00 & 94.79$\pm$0.33 & NIRC2 \\ 
2003.300 & 351.19$\pm$2.57 & 94.04$\pm$0.42 & SCAM  \\ 
2003.382 & 336.97$\pm$0.73 & 93.28$\pm$0.12 & FGS   \\ 
2003.514 & 340.06$\pm$0.61 & 93.99$\pm$0.10 & FGS   \\ 
2004.256 & 340.15$\pm$0.64 & 92.93$\pm$0.11 & NIRI  \\ 
2004.269 & 340.20$\pm$0.50 & 93.00$\pm$0.08 & NIRI  \\ 
2004.428 & 333.33$\pm$0.51 & 93.81$\pm$0.09 & FGS   \\ 
2004.534 & 337.14$\pm$1.40 & 93.72$\pm$0.24 & NIRC2 \\ 
\enddata 
\label{tab.sepPA}
\end{deluxetable} 

\clearpage

\begin{deluxetable}{lccc} 
\tablewidth{0pt}
\tablecaption{FGS PMT Counts} 
\tablehead{\colhead{Date} & \multicolumn{3}{c}{PMT Counts (per 25msec)\tablenotemark{\dag}}}
\startdata 
\multicolumn{1}{c}{DF Tau} & A & B &  \\
 \noalign{\vskip .8ex}%
 \hline
 \noalign{\vskip .8ex}%
2003/01/23 & 1285.1 &  668.2 &   \\
2003/11/19 & 1018.7 &  648.7 &   \\
 \noalign{\vskip 1.5ex}%
 \hline
 \noalign{\vskip .8ex}%
\multicolumn{1}{c}{ZZ Tau} & A & B &  \\
 \noalign{\vskip .8ex}%
 \hline
 \noalign{\vskip .8ex}%
2003/09/21 &  132.1 &   43.8 &   \\
2005/11/06 &  119.2 &   43.7 &   \\
 \noalign{\vskip 1.5ex}%
 \hline
 \noalign{\vskip .8ex}%
\multicolumn{1}{c}{V853 Oph} & A & B &  \\
 \noalign{\vskip .8ex}%
 \hline
 \noalign{\vskip .8ex}%
  1999/6/10 &  279.9 &   74.7 &  \\
  2000/4/24 &  233.1 &   60.7 &  \\
  2000/6/25 &  194.6 &   79.1 &  \\
  2001/5/01 &  185.6 &   65.1 &  \\
  2001/6/18 &  229.7 &   79.1 &  \\
  2001/7/04 &  193.9 &   71.0 &  \\
  2002/5/04 &  260.3 &   83.4 &  \\
  2003/5/20 &  287.0 &   65.9 &  \\
  2003/7/07 &  291.8 &   76.5 &  \\
  2004/6/05 &  207.0 &   58.3 &  \\
 \noalign{\vskip 1.5ex}%
 \hline
 \noalign{\vskip .8ex}%
\multicolumn{1}{c}{Elias 12} & S & Na & Nb \\
 \noalign{\vskip .8ex}%
 \hline
 \noalign{\vskip .8ex}%
1999/10/03  & 1841.2 &  261.3 &  153.9 \\
2000/08/27  & 1848.8 &  273.6 &  144.2 \\
2000/09/04  & 1939.1 &  245.7 &  117.7 \\
2000/12/31  & 1899.1 &  259.1 &   87.8 \\
2001/03/04  & 1937.0 &  258.4 &   81.6 \\
2002/12/22  & 1877.1 &  254.3 &  166.7 \\
2002/12/26  & 1585.7 &  258.7 &  153.7 \\
2003/11/08  & 1970.0 &  258.4 &  120.7 \\
2003/12/23  & 2043.0 &  227.5 &  127.1 \\
2005/08/06  & 1904.8 &  259.4 &  135.6 \\
\enddata
\tablenotetext{\dag}{Uncertainties are dominated by the square root of the number of counts derived from counting statistics.}
\label{tab.fgscounts} 
\end{deluxetable} 

\clearpage

\begin{deluxetable}{lccc} 
\tablewidth{0pt}
\tablecaption{Adaptive Optics Flux Ratios} 
\tablehead{\colhead{Date} & \colhead{Filter} & \colhead{Close Pair\tablenotemark{\dag}} & \colhead{Triple Flux Ratio\tablenotemark{\dag}}}
\startdata 
\multicolumn{2}{c}{DF Tau} & B/A &   \\
 \noalign{\vskip .8ex}%
 \hline
 \noalign{\vskip .8ex}%
2004/12/24 &    Kcont  &  0.475 &  \nodata \\
2005/12/06 &    K$'$   &  0.457 &  \nodata \\
2005/12/06 &    H      &  0.550 &  \nodata \\
 \noalign{\vskip 1.5ex}%
 \hline
 \noalign{\vskip .8ex}%
\multicolumn{2}{c}{ZZ Tau} & B/A &   \\
 \noalign{\vskip .8ex}%
 \hline
 \noalign{\vskip .8ex}%
2004/12/24 &  K$'$     &  0.672 &  \nodata \\
  \noalign{\vskip 1.5ex}%
 \hline
 \noalign{\vskip .8ex}%
\multicolumn{2}{c}{V853 Oph} & B/A &   \\
 \noalign{\vskip .8ex}%
 \hline
 \noalign{\vskip .8ex}%
2003/04/15 &     H      & 0.340 &  \nodata \\
2003/04/15 &     K      & 0.287 &  \nodata \\
2003/04/20 &  K$'$      & 0.270 &  \nodata \\
2004/04/03 & Br$\gamma$ & 0.227 &  \nodata \\
2004/04/03 &  FeII      & 0.275 &  \nodata \\
2004/04/08 & Br$\gamma$ & 0.197 &  \nodata \\
2004/04/08 &  FeII      & 0.235 &  \nodata \\
2004/07/14 &     H      & 0.354 &  \nodata \\
2004/07/14 &  K$'$      & 0.297 &  \nodata \\
  \noalign{\vskip 1.5ex}%
 \hline
 \noalign{\vskip .8ex}%
\multicolumn{2}{c}{Elias 12}  &  Nb/Na  &  S : Na : Nb  \\
 \noalign{\vskip .8ex}%
 \hline
 \noalign{\vskip .8ex}%
2002/03/08 &     H & 0.720 & 1.0 : 0.388 : 0.279 \\
2002/03/08 &  K$'$ & 0.692 & 1.0 : 0.399 : 0.276 \\
2002/10/30 &     H & 0.698 & 1.0 : 0.381 : 0.266 \\
2002/10/30 &  K$'$ & 0.723 & 1.0 : 0.322 : 0.233 \\
2004/12/24 &     H & 0.784 & 1.0 : 0.710 : 0.556 \\
2004/12/24 &  K$'$ & 0.748 & 1.0 : 0.606 : 0.452 \\
2005/03/01 & Hcont & 0.689 & 1.0 : 0.393 : 0.271 \\
2005/03/01 & Kcont & 0.731 & 1.0 : 0.308 : 0.224 \\
2005/12/06 &     H & 0.731 & 1.0 : 0.388 : 0.284 \\
2005/12/06 &  K$'$ & 0.736 & 1.0 : 0.326 : 0.240 \\
 \noalign{\vskip 1.5ex}%
 \hline
 \noalign{\vskip .8ex}%
\multicolumn{2}{c}{T Tau}   & Nb/Na  &  N : Sa : Sb  \\
 \noalign{\vskip .8ex}%
 \hline
 \noalign{\vskip .8ex}%
2002/10/30 & Br$\gamma$ & 3.45 & 1.0 : 0.020 : 0.069 \\
2002/10/30 &  K$'$      & 3.38 & 1.0 : 0.022 : 0.074 \\
2003/11/18 &  Kcont     & 5.14 & 1.0 : 0.011 : 0.054 \\
2004/12/24 &  Kcont     & 2.99 & 1.0 : 0.022 : 0.065 \\
2005/02/09 &  Kcont     & 8.00 & 1.0 : 0.009 : 0.071 \\
2005/03/09 &  Kcont     & 4.76 & 1.0 : 0.010 : 0.049 \\
2005/03/24 &    H2      & 3.71 & 1.0 : 0.014 : 0.052 \\
2005/03/24 &  Kcont     & 4.07 & 1.0 : 0.012 : 0.050 \\
2005/10/20 &  Kcont     & 2.87 & 1.0 : 0.020 : 0.058 \\
2005/12/06 &  K$'$      & 1.95 & 1.0 : 0.035 : 0.067 \\
\enddata
\tablenotetext{\dag}{Typical uncertainties for the AO flux ratios are $\pm 0.022$ for DF Tau B/A, $\pm 0.047$ for ZZ Tau B/A, $\pm$0.006 for V853 Oph B/A, $\pm$0.019 for Elias 12 Na/S, $\pm$0.012 for Elias 12 Nb/S, $\pm$0.001 for T Tau Sa/N, and $\pm$0.002 for T Tau Sb/N.}
\label{tab.aoflux} 
\end{deluxetable} 

\clearpage

\begin{deluxetable}{lcrr} 
\tablewidth{0pt}
\tablecaption{T Tau Component Magnitudes (assumes a magnitude of K=5.53 for T Tau N)} 
\tablehead{\colhead{Date} & \colhead{Filter} & \colhead{Sa} & \colhead{Sb}}
\startdata 
2002/10/30 & Br$\gamma$ &  9.78$\pm$0.04 & 8.44$\pm$0.04 \\
2002/10/30 &  K$'$      &  9.67$\pm$0.06 & 8.35$\pm$0.05 \\
2003/11/18 &  Kcont     & 10.47$\pm$0.09 & 8.69$\pm$0.04 \\
2004/12/24 &  Kcont     &  9.69$\pm$0.04 & 8.50$\pm$0.05 \\
2005/02/09 &  Kcont     & 10.66$\pm$0.04 & 8.40$\pm$0.03 \\
2005/03/09 &  Kcont     & 10.50$\pm$0.06 & 8.81$\pm$0.04 \\
2005/03/24 &    H2      & 10.16$\pm$0.04 & 8.73$\pm$0.04 \\
2005/03/24 &  Kcont     & 10.30$\pm$0.04 & 8.78$\pm$0.05 \\
2005/10/20 &  Kcont     &  9.76$\pm$0.04 & 8.62$\pm$0.03 \\
2005/12/06 &  K$'$      &  9.18$\pm$0.10 & 8.47$\pm$0.07 \\
\enddata
\label{tab.ttaumag} 
\end{deluxetable}

\clearpage

\begin{deluxetable}{ccccc}
\tablewidth{0pt}
\tablecaption{Orbital parameters from the 1-$\sigma$ confidence intervals and dynamical masses from the weighted Monte Carlo distributions}
\tablehead{\colhead{Parameter} & \colhead{DF Tau} & \colhead{ZZ Tau} & \colhead{Elias 12 Na-Nb} & \colhead{T Tau Sa-Sb}}
\startdata
$P$ (yr)               & $>$37     & $>$31     & 9.4$^{+1.2}_{-0.8}$       & $>$15     \\
$T$                    & 1974-2022 & 1994-1999 & 1991.16$^{+0.17}_{-0.23}$ & 1991-2001 \\
$e$                    & 0.05-0.76 & 0.35-0.88 & 0.47$^{+0.10}_{-0.11}$    & 0.09-0.87 \\
$a$ (mas)              & $>$96     & $>$60     & 36.8$^{+2.8}_{-1.7}$      & $>$82     \\
$i$ ($^{\circ}$)       & 128-176   & 115-134   & 147$^{+20}_{-11}$         & 11-60     \\
$\Omega (^{\circ})$  & \nodata   & 111-129   & 97$^{+20}_{-16}$          & \nodata   \\
$\omega (^{\circ})$  & \nodata   & 268-336   & 114$^{+29}_{-7.6}$        & \nodata   \\
 \noalign{\vskip .8ex}%
 \hline
 \noalign{\vskip .8ex}%
$M(\frac{d}{140 pc})^3$ ($M_{\odot}$)      & 0.78$^{+0.25}_{-0.15}$ & 0.66$^{+0.15}_{-0.11}$ & 1.13$^{+0.36}_{-0.09}$ & 4.13$^{+1.58}_{-0.97}$ \\
\enddata
\label{tab.orbpar}
\end{deluxetable}

\begin{deluxetable}{cc}
\tablewidth{0pt}
\tablecaption{Orbital Parameters for ZZ Tau A-B}
\tablehead{\colhead{Parameter} & \colhead{AO+FGS+LO} }
\startdata
$P$ (yr)               &  31.5$\pm$5.3     \\
$T$                    &  1994.26$\pm$0.85    \\
$e$                    &  0.401$\pm$0.085   \\
$a$ (mas)              &  61.$\pm$10.     \\
$i$ ($^{\circ}$)       &  133.9$\pm$7.6     \\
$\Omega (^{\circ})$  &  127.1$\pm$7.6     \\
$\omega (^{\circ})$  &  270.$\pm$20.     \\
$M(\frac{d}{140 pc})^3$ ($M_{\odot}$)      &  0.64$\pm$0.38           \\
$\chi^2$	     &  20.62  \\
$\nu$  		     &  20  \\
\enddata
\label{tab.zztauorb}
\end{deluxetable}

\clearpage

\begin{deluxetable}{cccc}
\tablewidth{0pt}
\tablecaption{Orbital Parameters for Elias 12 Na-Nb}
\tablehead{\colhead{Parameter} & \colhead{AO+FGS} & \colhead{AO+FGS+LO} & \colhead{AO+FGS+LO}}
\startdata
$P$ (yr)               &  9.4$\pm$1.0      & 8.65$\pm$0.14    & 11.85$\pm$0.12   \\
$T$                    &  1999.16$\pm$0.21 & 1999.29$\pm$0.10 & 1999.00$\pm$0.24 \\
$e$                    &  0.48$\pm$0.11    & 0.570$\pm$0.031  & 0.300$\pm$0.035  \\
$a$ (mas)              &  36.7$\pm$2.3     & 39.1$\pm$1.0     & 37.41$\pm$0.78   \\
$i$ ($^{\circ}$)       &  147.$\pm$14.     & 137.2$\pm$2.8    & 158.4$\pm$5.0    \\
$\Omega (^{\circ})$  &  97.$\pm$11.      & 97.4$\pm$5.7     & -1.$\pm$27.      \\
$\omega (^{\circ})$  &  115.$\pm$13.     & 108.4$\pm$4.1    & 39.$\pm$31.      \\
$M(\frac{d}{140 pc})^3$ ($M_{\odot}$)      &  1.53$\pm$0.45    & 2.19$\pm$0.19    & 1.026$\pm$0.068 \\
$\chi^2$	     &  8.46             & 8.78             & 9.43 \\
$\nu$  		     &  23               & 24               & 24 \\
\enddata
\label{tab.el12orb}
\end{deluxetable}

\clearpage

\clearpage

\clearpage	

\clearpage

\clearpage

\clearpage

\clearpage

\clearpage

\clearpage

\clearpage

\clearpage

\clearpage

\clearpage

\clearpage

\clearpage

\end{document}